\documentclass[reqno]{amsart}

\usepackage{amsthm,amsmath,amsfonts,amssymb,epsfig,graphicx,latexsym,float,color}
\usepackage{pgfplots}
\usetikzlibrary{plotmarks}

\newtheorem{definition}{Definition}

\newtheorem{remark}[definition]{Remark}

\newtheorem{system}[definition]{System}

\renewcommand{\theequation}{\arabic{section}.\arabic{equation}}

\renewcommand{\thefigure}{\arabic{section}.\arabic{figure}}

\begin{document}

\title[Melt-Blowing of Viscoelastic Jets in Turbulent Airflows]{Melt-Blowing of Viscoelastic Jets in Turbulent Airflows: Stochastic Modeling and Simulation}

\author[Wieland et al.]{Manuel Wieland$^{1}$}
\author[]{Walter Arne$^{1}$}
\author[]{Nicole Marheineke$^{2}$}
\author[]{Raimund Wegener$^{1}$}

\date{\today\\
$^1$ Fraunhofer ITWM, Fraunhofer Platz 1, D-67663 Kaiserslautern, Germany\\
$^2$ Universit\"at Trier, Lehrstuhl Modellierung und Numerik, Universit\"atsring 15, D-54296 Trier, Germany}

\begin{abstract}
In melt-blowing processes micro- and nanofibers are produced by the extrusion of polymeric jets into a directed, turbulent high-speed airflow. Up to now the physical mechanism for the drastic jet thinning is not fully understood, since in the existing literature the numerically computed/predicted fiber thickness differs several orders of magnitude from those experimentally measured. Recent works suggest that this discrepancy might arise from the neglect of the turbulent aerodynamic fluctuations in the simulations. In this paper we confirm this suggestion numerically. Due to the complexity of the process direct numerical simulations of the multiscale-multiphase problem are not possible. Hence, we develop a numerical framework for a growing fiber in turbulent air that makes the simulation of industrial setups feasible. For this purpose we employ an asymptotic viscoelastic model for the fiber. The turbulent effects are taken into account by a stochastic aerodynamic force model where the underlying velocity fluctuations are reconstructed from a $k$-$\epsilon$ turbulence description of the airflow. Our numerical results show the significance of the turbulence on the jet thinning and give fiber diameters of realistic order of magnitude.
\end{abstract}
\maketitle
{\sc AMS-Classification.} 35Lxx, 68U20, 76-XX
%
\setcounter{equation}{0} \setcounter{figure}{0} \setcounter{table}{0}
\section{Introduction}
\begin{figure}[!t]
\centering
\includegraphics[height=5cm]{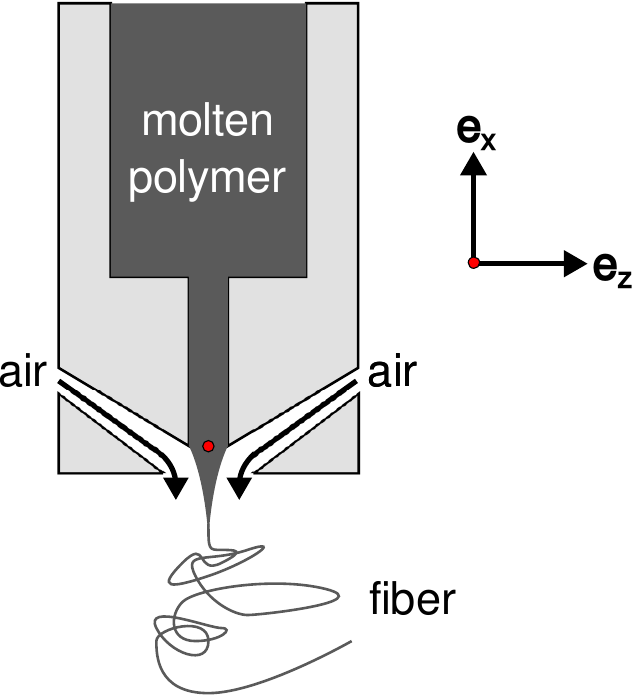}
\caption{Sketch of a typical melt-blowing setup.
}\label{sec:intro_fig:apparatus}
\end{figure}

Melt-blowing is a widely used production method for polymer micro- and nanofibers economically attractive due to low production costs. Fabrics of meltblown fibers are nonwovens, e.g., filters, hygiene products, battery separators. Details on the technology can be found in \cite{dutton:p:2009, pinchuk:b:2002}. A typical setup of a melt-blowing device is illustrated in Fig.~\ref{sec:intro_fig:apparatus}. In the process, molten polymer is fed through a nozzle into a forwarding high-speed and highly turbulent air stream to be stretched and cooled down. The resulting fibers are laid down onto some collector, e.g., conveyor belt. In contrast to melt-spinning processes, where the stretching is caused by a mechanical take-up, in melt-blowing the fiber jet thinning is due to the driving high-velocity air stream with its turbulent nature. 

To deepen the understanding on the mechanism of jet thinning in melt-blowing extensive diverse studies have been performed in the last years, covering experimental investigations, e.g., \cite{bansal:p:1998, bresee:p:2003, ellison:p:2007, wu:p:1992, xie:p:2012}, combined experimental numerical works, e.g., \cite{yarin:p:2010, uyttendaele:p:1990, yarin:p:2010b}, as well as numerical computations, e.g., \cite{chen:p:2003, shambaugh:p:2011, sun:p:2011, zeng:p:2011}.
However, so far, there is an obvious gap between the experimental and numerical results for the achieved fiber thickness in literature. The existing numerical simulations underestimate the fiber elongation by several orders of magnitude, cf.\ \cite{xie:p:2012, uyttendaele:p:1990, shambaugh:p:2011, zeng:p:2011}. While experimental studies show fiber elongations $e \sim \mathcal{O}(10^6)$, $e = A_{in}/A = d_{in}^2/d^2$, meaning a reduction of $10^3$ in diameter $d$ and of $10^6$ in cross-sectional area $A$ compared to the values at the nozzle (indicated by the index $_{in}$), simulated elongations are of order  $e\sim\mathcal{O}(10^4)$. This is likely due to steady considerations and the neglect of turbulent aerodynamic effects \cite{chen:p:2003, chen:p:2005, shambaugh:p:2011}. Assuming an incompressible steady fiber jet the relation $uA = u_{in}A_{in}$ with scalar jet speed $u$ holds true. Hence, the computed elongation is restricted by the velocity $\mathbf{v}_\star$ of the surrounding air stream, i.e., $e = u/u_{in} < \lVert \mathbf{v}_\star \rVert_\infty/u_{in}$. This estimate turns out to be valid also in (instationary) melt-blowing simulations where the surrounding airflow is computed (even) on basis of a turbulence model when only mean airflow informations are taken into account in the aerodynamic driving of the fiber jet \cite{sun:p:2011, zeng:p:2011}. Experiments in \cite{yarin:p:2010, xie:p:2012} indicate the relevance of the turbulent effects for the jet thinning. In \cite{yarin:p:2010b} a viscoelastic fiber model based on an upper convected Maxwell description (UCM) has been employed for melt-blowing, which is opposed to random pulsations. This is done by applying perturbation frequencies on a rectilinear fiber jet leading to bending instabilities and causing significant stretching and thinning of the jet. The examination has been extended to multiple fibers, focusing on the prediction of fiber deposition patterns and fiber-size distributions in the resulting nonwovens in \cite{yarin:p:2011}. Latest works deal with the numerical investigation of the angular fiber distribution, the effect of uptake velocity as well as the lay-down on a rotating drum \cite{sinharay:p:2013, ghosal:p:2016, ghosal:p:2016b}. In \cite{huebsch:p:2013} the significance of turbulence for melt-blowing has been approached by studying the effect of turbulent aerodynamic velocity fluctuations on a simplified fiber model of ordinary differential equations. There, a $k$-$\epsilon$ turbulence description of the high-speed airflow serves as basis for the reconstruction of the velocity fluctuations, yielding a stochastic aerodynamic force acting on the fiber jet.  

The aim of this paper is to establish a numerical framework for fibers in turbulent air that makes the simulation of industrial melt-blowing processes feasible. For this purpose we bring together the two described approaches: we extend the random field sampling of \cite{huebsch:p:2013} to the instationary viscoelastic UCM fiber model of \cite{yarin:p:2010b}. Since the aerodynamic forces are the key player for the fiber behavior, we employ a one-way coupling of the outer air stream with the fibers by the help of the force model given in \cite{marheineke:p:2009b}. Of importance is the efficient and robust realization that enables us presenting numerical results of an industrial setup with an appropriate viscoelastic description of the fiber, the inclusion of temperature effects and the direct incorporation of the turbulence structure of the outer air stream for the first time in literature.

Regarding the viscoelastic UCM fiber model of \cite{yarin:p:2010b}, that is asymptotically derived by slender body theory in \cite{marheineke:p:2016}, in Lagrange description it can uniquely be written as quasilinear hyperbolic first order system of partial differential equations on a growing space-time domain. Its classification with respect to the growing fiber domain gives requirements on boundary conditions with regard to well-posedness of the mathematical problem formulation and suggests a parameterization of the fiber tangent by the help of spherical coordinates. The effects of turbulent fluctuations are calculated by the turbulence reconstruction procedure described in \cite{huebsch:p:2013} and coupled into the fiber model by an air force function. The resulting instationary problem is solved using finite volumes in space with numerical fluxes of Lax-Friedrichs type as well as employing the implicit Euler method in time. For an industrial melt-blowing setup we show the applicability of our model and numerical solution framework and demonstrate the relevance of the turbulent fluctuations causing fiber elongations of the expected higher order of magnitude compared to stationary simulations. From the repeated random sampling of fibers in the sense of the Monte Carlo method a distribution of the final fiber diameters is obtained that yields fiber diameters of realistic order of magnitude.

The paper is structured as follows. In Sec.~\ref{sec:model} we start with the instationary viscoelastic UCM fiber model, regarding its classification and correct closing with boundary conditions. Furthermore, we give a short survey in reconstructing the turbulent fluctuations of an underlying air stream. After that we discuss our numerical solution framework and the handling of the growing fiber domain in Sec.~\ref{sec:numerics}. In Sec.~\ref{sec:example} we consider an industrial melt-blowing setup, for which we present simulation results covering the turbulent effects due to the high-speed air stream.

%
\setcounter{equation}{0} \setcounter{figure}{0} \setcounter{table}{0}
\section{Viscoelastic fiber melt-blowing model}\label{sec:model}

For the melt-blowing of a fiber in a turbulent air stream we present an asymptotic instationary viscoelastic UCM fiber model in Lagrangian (material) description. We classify the resulting quasilinear system of partial differential equations of first order and discuss the appropriate closing by boundary conditions. The choice of the boundary conditions suggests a description with respect to a fiber tangent associated basis. The fiber tangent $\boldsymbol{\tau}$ with norm $e = \lVert \boldsymbol{\tau} \rVert$ and direction $\mathbf{t} = \boldsymbol{\tau}/\lVert \boldsymbol{\tau} \rVert$ is in particular parameterized by the help of spherical coordinates. Moreover, we present the models for the aerodynamic force and the heat exchange used in the one-way coupling with the surrounding airflow and introduce the stochastic modeling concept by which the effects of the turbulent aerodynamic velocity fluctuations are incorporated in the fiber system.

\subsection{Asymptotic fiber jet model}\label{sec:model_general}
The extrusion of a fiber jet from a nozzle into an air stream can be seen as an inflow problem with a domain enlarging over time. Let $\Omega= \{(\zeta,t)\in\mathbb{R}^2 \, \lvert \, \zeta\in\mathcal{Q}(t),\, t\in(0,t_{end}]\}$ be the space-time domain with time-dependent growing space $\mathcal{Q}(t) = (-\zeta_L(t),0)$, where $\mathrm{d}/\mathrm{d}t\, \zeta_L(t) = v_{in}(t)$, $\zeta_L(0) = 0$, with $v_{in}$ [m/s] being the (scalar) inflow velocity at the nozzle. In the following we assume a constant inflow velocity, i.e., $v_{in} = const$, yielding $\zeta_L(t) = v_{in}\, t$. The fiber jet is represented by a time-dependent curve $\mathbf{r}:\Omega\rightarrow \mathbb{R}^3$, where the fiber end corresponds to the material parameter $\zeta = 0$ and the material points entering the fiber (flow) domain at the nozzle are $\zeta = -\zeta_L(t)$. We assume incompressibility of the fiber jet such that besides the mass also the volume is conserved, i.e., $\partial_t\varrho_M = 0$ and $\partial_t\varrho_V = 0$ with mass and volume line densities, $\varrho_M$ [kg/m] and $\varrho_V$ [m$^2$], respectively. The mass and volume line densities are considered to be constant at the nozzle yielding $\varrho_M = \rho d_{in}^2\pi/4$ and $\varrho_V = \lVert \boldsymbol{\tau}_{in} \rVert d_{in}^2\pi/4$ with constant fiber density $\rho$ [kg/m$^3$], nozzle diameter $d_{in}$ [m] and fiber tangent at the nozzle $\boldsymbol{\tau}_{in}$.  According to \cite{marheineke:p:2016}, where the viscoelastic UCM string model has been systematically derived by slender body asymptotics, our model for the extruding fiber jet is given in Lagrangian description by
\begin{align*}
\partial_t \mathbf{r} &= \mathbf{v},\\
\partial_\zeta \mathbf{r} &= \boldsymbol{\tau},\\
\partial_t(\varrho_M\mathbf{v}) &= \partial_\zeta\left(\varrho_V\sigma\frac{\boldsymbol{\tau}}{\lVert\boldsymbol{\tau}\rVert^2}\right) + \mathbf{f}_g + \mathbf{f}_{air},\\
c_p\partial_t(\varrho_M T) &= - \pi d \alpha (T-T_\star) \lVert \boldsymbol{\tau} \rVert,\\
\partial_t \sigma &= \left(3p+2\sigma+3\frac{\mu}{\theta}\right)\frac{\partial_t\lVert\boldsymbol{\tau}\rVert}{\lVert\boldsymbol{\tau}\rVert} - \frac{\sigma}{\theta},\\
\partial_t p &= \left(-p-\frac{\mu}{\theta}\right)\frac{\partial_t\lVert\boldsymbol{\tau}\rVert}{\lVert\boldsymbol{\tau}\rVert} - \frac{p}{\theta},
\end{align*}
supplemented with appropriate initial and boundary conditions to be specified. The diameter function $d:\mathcal{Q}\rightarrow\mathbb{R}^+$ is introduced via
\begin{align*}
d = 2\sqrt{\frac{\varrho_V}{\pi\lVert \boldsymbol{\tau} \rVert}}.
\end{align*}
The two kinematic equations relate the fiber velocity $\mathbf{v}$ [m/s] and the fiber tangent $\boldsymbol{\tau}$ to the derivatives of the fiber curve $\mathbf{r}$ [m] with respect to time $t$ [s] and material parameter $\zeta$ [m]. The two dynamic equations prescribing the conservation of linear momentum and energy yield equations for fiber velocity $\mathbf{v}$ and fiber temperature $T$ [K]. The acting outer line force densities arise from gravity $\mathbf{f}_g = \varrho_M g \mathbf{e}_g$ [N/m] with direction $\mathbf{e}_g$, $\|\mathbf{e}_g\|=1$, and gravitational constant $g$ [m/s$^2$] as well as from the surrounding airflow $\mathbf{f}_{air}$ [N/m]. Moreover, $\alpha$ [W/(m$^2$K)] is the heat transfer coefficient, $T_\star$ [K] the aerodynamic temperature field, $c_p$ [J/(kgK)] the constant specific heat capacity of the fiber and $d$ [m] the fiber diameter. The models for the aerodynamic line force density $\mathbf{f}_{air}$ and for the heat transfer coefficient $\alpha$ are presented in Sec.~\ref{subsec:drag_turb_nusselt}. Concerning the viscoelastic material laws, they are based on a UCM model for the fiber stress $\sigma$ [Pa] and pressure $p$ [Pa]. Here, $\mu$ [Pa s] describes the dynamic viscosity and $\theta$ [s] the relaxation time of the fiber jet. Under the assumption of incompressibility the relation $\mu/\theta = E/3$ with elastic modulus $E$ [Pa]  holds. We model the dynamic viscosity and relaxation time dependent on the temperature $T$, i.e., $\mu = \mu(T)$, $\theta = \theta(T)$. The corresponding rheological laws for an industrial example are specified in Sec.~\ref{sec:ex_subsec:setup}. 

For the numerical treatment of the problem it is convenient to deal with dimensionless model equations. We introduce the dimensionless quantities as $\tilde{y}(\tilde{\zeta},\tilde{t}) = y(\zeta_0\tilde{\zeta},t_0\tilde{t})/y_0$ and use the reference values $y_0$ as given in Tab.~\ref{sec:model_table:entdim}. Here, $y_{in}$ indicates the value of a quantity $y$ at the nozzle and $H$ denotes the height of the considered melt-blowing device. The constant mass and volume line densities $\varrho_M$, $\varrho_V$ become $\tilde{\varrho}_M = \tilde{\varrho}_V = 1$ in dimensionless form. To keep the notation simple we suppress the label  $\tilde{~}$ in the following. Then, the dimensionless model equations read
\begin{equation}\label{sec:model_eq:dimlessSystem}
\begin{aligned}
\partial_t \mathbf{r} &= \mathbf{v},\\
\partial_\zeta \mathbf{r} &= \boldsymbol{\tau},\\
\partial_t \mathbf{v} &= \partial_\zeta\left(\sigma\frac{\boldsymbol{\tau}}{\lVert\boldsymbol{\tau}\rVert^2}\right) + \frac{1}{\mathrm{Fr}^2}\mathbf{e}_g + \mathbf{f}_{air},\\
\partial_t T &= -\frac{\mathrm{St}}{\varepsilon}\pi d\alpha(T-T_\star)\lVert\boldsymbol{\tau} \rVert,\\
\mathrm{De}\left(\partial_t \sigma - (2\sigma+3p) \frac{\partial_t\lVert\boldsymbol{\tau}\rVert}{\lVert\boldsymbol{\tau}\rVert}\right) +\frac{\sigma}{\theta} &= \frac{3}{\mathrm{Re}}\frac{\mu}{\theta}\frac{\partial_t\lVert\boldsymbol{\tau}\rVert}{\lVert\boldsymbol{\tau}\rVert},\\
\mathrm{De}\left(\partial_t p + p\frac{\partial_t\lVert\boldsymbol{\tau}\rVert}{\lVert\boldsymbol{\tau}\rVert}\right) +\frac{p}{\theta} &= - \frac{1}{\mathrm{Re}}\frac{\mu}{\theta}\frac{\partial_t\lVert\boldsymbol{\tau}\rVert}{\lVert\boldsymbol{\tau}\rVert}.
\end{aligned}
\end{equation}
The fiber behavior is characterized by the dimensionless parameters given in Tab.~\ref{sec:model_table:entdim}, that are the Reynolds number $\mathrm{Re}$ as ratio of inertial to viscous forces, the Deborah number $\mathrm{De}$ as ratio of relaxation time to characteristic time-scale, the Froude number $\mathrm{Fr}$ as ratio of inertial to gravitational forces, the Stanton number $\mathrm{St}$ as ratio of heat transfer to thermal capacity as well as the slenderness ratio $\varepsilon$. The time-dependent space domain simplifies to $\mathcal{Q}(t) = (-t,0)$.

\begin{remark}\label{sec:model_remark:viscousLimit}
The viscoelastic UCM fiber model~\eqref{sec:model_eq:dimlessSystem} covers the limit cases describing pure viscous as well as elastic material behavior. The limit $\mathrm{De}\rightarrow 0$ yields a viscous fiber model, whereas the limit $\mathrm{Re}\rightarrow 0$, $\mathrm{De}\rightarrow\infty$ with $\mathrm{Re}\mathrm{De} = \mathrm{Ma}^2$ describes an elastic behavior. Here, the dimensionless Mach number $\mathrm{Ma}$ is the ratio of inertial to compressive forces.
\end{remark}

\begin{remark}\label{sec:model_remark:p}
As pointed out in \cite{marheineke:p:2016}, the pressure $p$ is at least one order of magnitude smaller than the stress $\sigma$ for fibers with high strain rates $\upsilon = \partial_t\lVert\boldsymbol{\tau}\rVert / \lVert\boldsymbol{\tau}\rVert \geq 0$ and large Deborah numbers $\mathrm{De}$, in particular $\lvert p \rvert \leq 0.1 \sigma$ if $\upsilon\mathrm{De}\,\theta \geq 0.35$. This means the pressure equation can be neglected in such cases. In \cite{yarin:p:2010b} this simplification is employed instantaneously to the UCM model for melt-blowing.
\end{remark}
\begin{table}[t]
\begin{minipage}[c]{\textwidth}
\begin{center}
\begin{small}
\begin{tabular}{| l r@{ = } l l |}
\hline
\multicolumn{4}{|l|}{\textbf{Reference values}}\\
Description & \multicolumn{2}{l}{Formula} & Unit\\
\hline
fiber curve & $r_0$ & $H$ & m \rule{0pt}{2.6ex}\\
fiber diameter & $d_0$ & $d_{in}\sqrt{\pi}/2$ & m\\
fiber velocity & $v_0$ & $v_{in}$ & m/s \\
fiber temperature & $T_0$ & $T_{in}$ & K\\
fiber mass line density & $\varrho_{M0}$ & $\rho d_0^2$ & kg/m\\
fiber volume line density & $\varrho_{V0}$ & $d_0^2$ & m$^2$\\
fiber stress & $\sigma_0$ & $\varrho_{M0}v_0^2/d_0^2$ & Pa\\
fiber pressure & $p_0$ & $\sigma_0$ & Pa\\
fiber kinematic viscosity & $\mu_0$ & $\mu(T_0)$ & Pas\\
fiber relaxation time & $\theta_0$ & $\theta(T_0)$ & s\\
outer forces & $f_0$ & $\varrho_{M0}v_0^2/r_0$ & N/m\\
heat transfer coefficient & $\alpha_0$ & $\alpha_{in}$ & W/(m$^2$K)\\
length scale & $\zeta_0$ & $r_0$ & m\\
time scale & $t_0$ & $r_0/v_0$ & s\\
air velocity & $v_{\star,0}$ & $v_0$ & m/s\\
air density & $\rho_{\star,0}$ & $\rho_{\star,in}$ & kg/m$^3$\\
air kinematic viscosity & $\nu_{\star,0}$ & $\nu_{\star,in}$ & m$^2$/s\\
air specific heat capacity & $c_{p,\star,0}$ & $c_{p,\star,in}$ & J/(kgK)\\
air thermal conductivity & $\lambda_{\star,0}$ & $\lambda_{\star,in}$ & W/(mK)\\
air turbulent kinetic energy & $k_{\star,0}$ & $k_{\star,in}$ & m$^2$/s$^2$\\
air viscous dissipation rate & $\epsilon_{\star,0}$ & $\epsilon_{\star,in}$ & m$^2$/s$^3$\\
\hline
\end{tabular}
\end{small}
\end{center}
\end{minipage}\vfill
\vspace*{0.5cm}
\begin{minipage}[c]{\textwidth}
\begin{center}
\begin{small}
\begin{tabular}{| l r@{ = } l |}
\hline
\multicolumn{3}{|l|}{\textbf{Dimensionless numbers}}\\
Description & \multicolumn{2}{l|}{Formula}\\
\hline
slenderness & $\varepsilon$ & $d_0/r_0$ \rule{0pt}{2.6ex}\\
Reynolds & $\mathrm{Re}$ & $\varrho_{M0}v_0r_0/(d_0^2\mu_0)$ \\
Deborah & $\mathrm{De}$ & $\theta_0/t_0$\\
Froude & $\mathrm{Fr}$ & $v_0/\sqrt{gr_0}$ \\
Stanton & $\mathrm{St}$ & $d_0^2\alpha_0/(c_p\varrho_{M0}v_0)$\\
Mach & $\mathrm{Ma}$ & $v_0/d_0\sqrt{\varrho_{M0}\theta_0/\mu_0}$\\
air drag associated & $\mathrm{A}_\star$ & $\rho_{\star,0}d_0v_0^2/f_0$\\
mixed (air-fiber) Reynolds & $\mathrm{Re}_\star$ & $d_0v_0/\nu_{\star,0}$\\
Nusselt & $\mathrm{Nu}_\star$ & $\alpha_0d_0/\lambda_{\star,0}$\\
Prandtl & $\mathrm{Pr}_\star$ & $c_{p,\star,0}\rho_{\star,0}\nu_{\star,0}/\lambda_{\star,0}$\\
turbulence degree & $\mathrm{Tu}_\star$ & $k_{\star,0}^{1/2}/v_0$\\
turbulent time & $\mathrm{Tt}_\star$ & $\epsilon_{\star,0}r_0/(k_{\star,0}v_0)$\\
\hline
\end{tabular}
\end{small}
\end{center}
\end{minipage}\\~\\~\\
\caption{Overview over reference values used for non-dimensionalization and the resulting dimensionless numbers.}\label{sec:model_table:entdim}
\end{table}

\subsection{Classification and boundary conditions}
The dimensionless fiber model (\ref{sec:model_eq:dimlessSystem}) can uniquely be written as a quasilinear system of partial differential equations of first order \cite{marheineke:p:2016}
\begin{align}\label{sec:model_eq:quasilinearForm}
\partial_t\boldsymbol{\varphi} + \mathbf{M}(\boldsymbol{\varphi})\cdot\partial_\zeta\boldsymbol{\varphi} + \mathbf{m}(\boldsymbol{\varphi}) = \mathbf{0}
\end{align}
with the vector of unknowns $\boldsymbol{\varphi} = (\mathbf{r},\boldsymbol{\tau},\mathbf{v},T,\sigma,p)\in\mathbb{R}^{12}$. The system is classified mathematically by the spectrum of the system matrix $\mathbf{M}$ that consists of the eigenvalues
\begin{itemize}
\item $\lambda_1 = 0$ (multiplicity $6$),
\item $\lambda_{2,3} = \pm\sqrt{\sigma}/\lVert \boldsymbol{\tau} \rVert$ (multiplicity $2$ each),
\item $\lambda_{4,5} = \pm\sqrt{w}/\lVert \boldsymbol{\tau} \rVert$ (multiplicity $1$ each), $\qquad w = \left(3\mu/\theta+\mathrm{Ma}^2\left(\sigma + 3p\right)\right)/\mathrm{Ma}^2$
\end{itemize}
The system is of hyperbolic type if $\sigma > 0$ and $w >  0$. Otherwise, it is mixed elliptic-hyperbolic, or even shows a parabolic deficiency if $\sigma=0$ and/or $w=0$.

Since the hyperbolic case is relevant for the application, we focus on it and discuss the closing of the system by appropriate boundary and initial conditions. At the fiber jet end, which corresponds to a fixed material point in Lagrangian description ($\zeta = 0$), the characteristic related to the eigenvalue $\lambda_i$ runs from the nozzle to the jet end if $\lambda_i > 0$ and from the jet end towards the nozzle if $\lambda_i < 0$. At the nozzle ($\zeta = -\zeta_L(t)$) the orientations of the characteristics depend on the scalar inflow velocity of the fiber jet, which reads  $v_{in}/v_0 = 1$ in non-dimensional form. If $\lambda_i  > -v_{in}/v_0=-1$ for $i\in\{1,...,5\}$, the corresponding characteristic propagates from the nozzle to the jet end, otherwise the other way round. The orientations of the characteristics yield requirements on the boundary conditions with regard to the well-posedness of the problem. Since $\lambda_3 < 0$ (multiplicity 2) and $\lambda_5 < 0$ (multiplicity 1), we have to pose three boundary conditions at the fiber jet end. Because of the spinning setup we model the fiber end ($\zeta = 0$) as stress-free, i.e.,
\begin{align*}
\sigma(0,t) = 0, \qquad p(0,t) = 0.
\end{align*}
Employing the viscoelastic material law for $\sigma$ yields a constant fiber elongation $e = \lVert \boldsymbol{\tau} \rVert$ at the fiber end over time, i.e., $\partial_t e(0,t) = 0$. To preserve this compatibility condition we pose
\begin{align*}
e(0,t) = 1,
\end{align*}
assuming the fiber jet to leave the nozzle unstretched. The eigenvalues $\lambda_1$, $\lambda_2$, $\lambda_4$ are non-negative and thus imply nine boundary conditions at the nozzle ($\zeta = -\zeta_L(t)$, $t\geq0$),
\begin{align*}
\mathbf{r}(-\zeta_L(t),t) &= \mathbf{r}_{in}/r_0, \qquad \left(\boldsymbol{\tau}/e\right)(-\zeta_L(t),t) = \mathbf{e}_g, \qquad
\mathbf{v}(-\zeta_L(t),t) = \mathbf{e}_g, \qquad T(-\zeta_L(t),t) = 1.
\end{align*}
Here, $\mathbf{r}_{in}$ is assumed to be constant. Furthermore, we set the following initial conditions for $t=0$,
\begin{align*}
\sigma(-\zeta_L(0),0) = \sigma_{in}/\sigma_0,\qquad 
p(-\zeta_L(0),0) = p_{in}/p_0, \qquad e(-\zeta_L(0),0) = 1.
\end{align*}
Depending on the propagation-speed of the characteristics at the nozzle we pose further boundary conditions: we additionally prescribe for $t>0$
\begin{align*}
 \sigma(-\zeta_L(t),t) &= \sigma_{in}/\sigma_0, \qquad p(-\zeta_L(t),t) = p_{in}/p_0,&& \text{ if } \lambda_3 > -1  \text{ (multiplicity 2)}, \\
e(-\zeta_L(t),t) &= 1 && 
\text{ if } \lambda_5 > -1  \text{ (multiplicity 1)}.
\end{align*}
The total time-derivative of the fiber curve $\mathbf{r}$ at the nozzle yields the compatibility condition $\mathbf{v}(-\zeta_L(t),t) = \boldsymbol{\tau}(-\zeta_L(t),t)$ for all times $t$. Through the above choice of the boundary conditions for $\mathbf{v}$, $\boldsymbol{\tau}/e$, and $e$ at the nozzle this condition is inherently fulfilled.

The choice of the boundary conditions and in particular the decomposition of the fiber tangent $\boldsymbol{\tau}=e\mathbf{t}$ into elongation $e = \lVert \boldsymbol{\tau}\rVert$ and direction $\mathbf{t}$, $\|\mathbf{t}\|=1$, suggests a reformulation of the corresponding dynamic equation $\partial_\zeta \mathbf{r}= \boldsymbol{\tau}$. Making use of the compatibility condition $\partial_t\boldsymbol{\tau} = \partial_t\partial_\zeta\mathbf{r} = \partial_\zeta\partial_t\mathbf{r} = \partial_\zeta \mathbf{v}$ yields an equation for the elongation $e$
\begin{align*}
\partial_t e - \mathbf{t}\cdot\partial_\zeta\mathbf{v} = 0.
\end{align*}
The normalized tangent $\mathbf{t}$ can be parameterized by means of spherical coordinates
\begin{align*}
\mathbf{t}(\vartheta,\varphi) = (\sin\vartheta\cos\varphi, \sin\vartheta\sin\varphi,\cos\vartheta), \qquad \vartheta \in [0,\pi], \quad \varphi \in [0,2\pi). 
\end{align*}
Then, its time-derivative reads
$ \partial_t\mathbf{t} = \mathbf{n}\partial_t\vartheta + \mathbf{b}\partial_t\varphi $
with normal $\mathbf{n} = (\cos\vartheta\cos\varphi, \cos\vartheta\sin\varphi, -\sin\vartheta)$ and binormal $\mathbf{b} = (-\sin\vartheta\sin\varphi, \sin\vartheta\cos\varphi, 0)$. The set $\{\mathbf{t},\mathbf{n},\mathbf{b}\} \subset \mathbb{R}^3$ forms an orthogonal basis where $\|\mathbf{t}\|=\|\mathbf{n}\|=1$. Employing
\begin{align*}
\partial_t\mathbf{t} = \partial_t\left(\frac{\boldsymbol{\tau}}{e}\right) = \frac{1}{e}\left(\mathbf{I} - \mathbf{t}\otimes\mathbf{t}\right)\cdot\partial_\zeta\mathbf{v}
\end{align*}
gives relations for the polar $\vartheta$ and azimuth angles $\varphi$
\begin{align*}
\partial_t\vartheta = \frac{1}{e}\mathbf{n}\cdot\partial_\zeta\mathbf{v},\qquad
\sin^2\vartheta\,\partial_t\varphi = \frac{1}{e}\mathbf{b}\cdot\partial_\zeta\mathbf{v}.
\end{align*}
Summing up, our viscoelastic instationary fiber model on a growing domain in Lagrangian description is given by System~\ref{system:Final}.
\begin{system}[Instationary viscoelastic fiber model]\label{system:Final}
Kinematic and dynamic equations as well as material laws in $\Omega$:
\begin{align*}
\partial_t \mathbf{r} - \mathbf{v}&=0,\\
\partial_t e - \mathbf{t}\cdot\partial_\zeta\mathbf{v} &= 0,\\
\partial_t\vartheta - \frac{1}{e}\mathbf{n}\cdot\partial_\zeta\mathbf{v} &= 0,\\
\sin^2\vartheta\,\partial_t\varphi - \frac{1}{e}\mathbf{b}\cdot\partial_\zeta\mathbf{v} &=0,\\
\partial_t\mathbf{v} - \partial_\zeta\left(\sigma\frac{\mathbf{t}}{e}\right) -  \frac{1}{\mathrm{Fr}^2}\mathbf{e}_g - \mathbf{f}_{air} &= 0,\\
\partial_t T + \frac{\mathrm{St}}{\varepsilon}\pi d\alpha(T-T_\star)e &= 0,\\
\mathrm{De}\,\partial_t \sigma + \left(-\mathrm{De}\,(2\sigma+3p)-\frac{\mu}{\theta}\frac{3}{\mathrm{Re}}\right) \,\frac{\mathbf{t}}{e}\cdot\partial_\zeta\mathbf{v}+ \frac{\sigma}{\theta} &= 0,\\
\mathrm{De}\,\partial_t p + \left(\mathrm{De}\,p+\frac{\mu}{\theta}\frac{1}{\mathrm{Re}}\right)\,\frac{\mathbf{t}}{e}\cdot\partial_\zeta\mathbf{v} + \frac{p}{\theta} &= 0,
\end{align*}
Initial-boundary conditions at the nozzle ($\zeta = -\zeta_L(t)$, $t \geq 0$):
\begin{align*}
\mathbf{r}(-\zeta_L(t),t) &= \mathbf{r}_{in}/r_0, \qquad &\vartheta(-\zeta_L(t),t) &= \vartheta_{in} \qquad &\varphi(-\zeta_L(t),t) &= \varphi_{in},\\
\mathbf{v}(-\zeta_L(t),t) &= \mathbf{e}_g, \qquad & T(-\zeta_L(t),t) &= 1,
\end{align*}
Initial conditions ($t=0$):
\begin{align*}
e(-\zeta_L(0),0) = 1,\qquad \sigma(-\zeta_L(0),0) = \sigma_{in}/\sigma_0,\qquad 
p(-\zeta_L(0),0) = p_{in}/p_0,
\end{align*}
Boundary conditions at the nozzle ($\zeta = -\zeta_L(t)$, $t > 0$):
\begin{align*}
&\text{if } \lambda_3>-1 \text{: }& \sigma(-\zeta_L(t),t) &= \sigma_{in}/\sigma_0, \qquad p(-\zeta_L(t),t) = p_{in}/p_0,\\
&\text{if } \lambda_5>-1 \text{: }& e(-\zeta_L(t),t) &= 1, 
\end{align*}
Boundary conditions at the fiber end ($\zeta = 0$, $t>0$):
\begin{align*}
e(0,t) = 1, \qquad \sigma(0,t) = 0, \qquad p(0,t) = 0.
\end{align*}
\end{system}

\subsection{Exchange models for one-way coupling with turbulent airflow}\label{subsec:drag_turb_nusselt}
In this work we consider a one-sided coupling of the airflow with the fiber, neglecting feedback effects of the fiber on the airflow. The respective exchange models used for the aerodynamic line force density $\mathbf{f}_{air}$ and the heat transfer coefficient $\alpha$ are briefly summarized in this subsection. Moreover, we describe the concept how the turbulent aerodynamic velocity fluctuations are realized with respect to an underlying (stochastic) airflow simulation and incorporated in our fiber model (System~\ref{system:Final}). 

Note that to distinguish the fiber quantities from the airflow quantities, all airflow associated fields are labeled with the index $_\star$ as before. In particular, $\mathbf{v}_\star$ denotes the velocity, $\rho_\star$ the density, $\nu_\star$ the kinematic viscosity, $c_{p,\star}$ the specific heat capacity, $\lambda_\star$ the thermal conductivity, $k_\star$ the turbulent kinetic energy and $\epsilon_\star$ the viscous dissipation of the turbulent motions per unit mass of the air. All these quantities are space- and time-dependent fields assumed to be dimensionless and known -- for example provided by an external computation. The corresponding reference values used for non-dimensionalization are denoted with the index $_0$ and given in Tab.~\ref{sec:model_table:entdim}.

\subsubsection{Aerodynamic force and heat transfer coefficient}\label{subsubsec:drag}
The models for the aerodynamic force and the heat transfer coefficient are determined by material and geometrical properties as well as the incident flow situation which can be prescribed by the fiber orientation (normalized tangent) $\mathbf{t}$ and the relative velocity between airflow and fiber $\mathbf{v}_\star -\mathbf{v}$.

The aerodynamic line force density $\mathbf{f}_{air}$ is modeled by means of a dimensionless drag function $\mathbf{F}:\mathrm{SO}(3)\times \mathbb{R}^3\rightarrow \mathbb{R}^3$  which depends on fiber tangent and relative velocity,
\begin{align}\label{sec:model_eq:airdrag}
\mathbf{f}_{air} = e\frac{\mathrm{A}_\star}{\mathrm{Re}_\star^2}\frac{\rho_\star\nu^2_\star}{d} \mathbf{F}\bigg(\mathbf{t},\mathrm{Re}_\star\frac{d}{\nu_\star}(\mathbf{v}_\star - \mathbf{v})\bigg),\qquad
\mathbf{F}(\mathbf{t},\mathbf{w}) = r_n(w_n)\mathbf{w_n} + r_t(w_n)\mathbf{w_t}.
\end{align}
The drag function can be particularly expressed in terms of the tangential $\mathbf{w_t} = (\mathbf{w}\cdot\mathbf{t})\mathbf{t}$ and normal relative velocity components $\mathbf{w_n}=\mathbf{w}-\mathbf{w_t}$, $w_n = \lVert \mathbf{w_n} \rVert$. The models used for the tangential and normal air resistance coefficients $r_t$, $r_n$ are taken from \cite{marheineke:p:2009b}, see Appendix \ref{appendix_AirDrag} for details. The occurring dimensionless numbers are the air drag associated number $\mathrm{A}_\star$ and the mixed (air-fiber) Reynolds number $\mathrm{Re}_\star$ (cf. Tab.~\ref{sec:model_table:entdim}). Concerning lift forces see Remark~\ref{rem:lift}.

The heat transfer coefficient $\alpha$ is modeled by a Nusselt number associated dimensionless function $\mathcal{N}:\mathbb{R}^3\rightarrow\mathbb{R}$ which depends on the tangential and absolute relative velocity and the Prandtl number, 
\begin{align}
\alpha = \frac{1}{\mathrm{Nu}_\star}\frac{\lambda_\star}{d}\mathcal{N}\left(\mathrm{Re}_\star\frac{d}{\nu_\star}(\mathbf{v_\star} - \mathbf{v})\cdot\mathbf{t}, \mathrm{Re}_\star\frac{d}{\nu_\star}\lVert \mathbf{v_\star} - \mathbf{v}\rVert, \mathrm{Pr}_\star\frac{c_{p,\star}\rho_\star\nu_\star}{\lambda_\star}\right).
\end{align}
For details on the used heuristic model for $\mathcal{N}$ we refer to Appendix~\ref{appendix_Nusselt}. The occurring dimensionless numbers are the Nusselt number $\mathrm{Nu}_\star$, the Prandtl number $\mathrm{Pr}_\star$ as well as the mixed (air-fiber) Reynolds number $\mathrm{Re}_\star$ (cf. Tab.~\ref{sec:model_table:entdim}).

\subsubsection{Turbulence reconstruction}\label{subsubsec:turbRecon}
A direct numerical simulation of the turbulent airflow in the application is not possible due to the required high resolution. Hence, a statistical turbulence description is used where the airflow velocity $\mathbf{v}_\star$ is assumed to consist of a mean (deterministic) part $\bar{\mathbf{v}}_\star$ and a fluctuating (stochastic) part $\mathbf{v}'_\star$, i.e.,
\begin{align*}
\mathbf{v}_\star = \bar{\mathbf{v}}_\star + \mathbf{v}'_\star.
\end{align*}
The mean velocity is given by the Reynolds-averaged Navier-Stokes equations, while the fluctuations are only characterized by certain quantities that the respective turbulence model provides. To obtain $\mathbf{v}'_\star$ explicitly as random field we apply a turbulence reconstruction that has been developed in  \cite{huebsch:p:2013} on basis of a $k_\star$-$\epsilon_\star$ turbulence model. Assuming given dimensionless space-time-dependent fields for the turbulent kinetic energy $k_\star$ and the viscous dissipation of the turbulent motions per unit mass $\epsilon_\star$, the general concept of the turbulence reconstruction is to model the local turbulent fluctuations as homogeneous, isotropic, incompressible Gaussian random fields in space and time, $\mathbf{v}_{\star,loc}' = \mathbf{v}_{\star,loc}'(\mathbf{x}, t; \nu_\star, \bar{\mathbf{v}}_\star)$, that depend parametrically on the kinematic viscosity and mean velocity of the airflow, as done in \cite{marheineke:p:2006, marheineke:p:2009b}. To form the large-scale structure of the global turbulence the local fluctuations fields are superposed based on a Global-from-Local assumption. The globalization strategy according to \cite{huebsch:p:2013} yields
\begin{align}\label{sec:model_eq:turbReconstr}
\mathbf{v}_\star' = \mathrm{Tu}_\star k_\star^{1/2}\mathbf{v}_{\star,loc}'\left(\frac{\mathrm{Tt}_\star}{\mathrm{Tu}_\star}\frac{\epsilon_\star}{k_\star^{3/2}}\mathbf{r}, \mathrm{Tt}_\star\frac{\epsilon_\star}{k_\star}t; \frac{\varepsilon}{\mathrm{Re}_\star}\frac{\mathrm{Tt_\star}}{\mathrm{Tu}_\star^2}\frac{\epsilon_\star}{k_\star^2} \nu_\star,\frac{1}{\mathrm{Tu}_\star} \frac{1}{k_\star^{1/2}} \bar{\mathbf{v}}_\star\right).
\end{align}
Besides the slenderness ratio $\varepsilon$ and the mixed Reynolds number $\mathrm{Re}_\star$, the occurring dimensionless numbers are the degree of turbulence $\mathrm{Tu}_\star$ and the turbulent time scale ratio $\mathrm{Tt}_\star$ as given in Tab.~\ref{sec:model_table:entdim}. 

Note that the occurring turbulent length and time scales give requirements on the spatial and temporal resolution in our numerical solution algorithm  (cf.\ Rem.~\ref{sec:numerics_remark:resolution} in Sec.~\ref{sec:numerics}). In particular, $l'_\star = \mathrm{Tu}_\star/\mathrm{Tt}_\star k_\star^{3/2}/\epsilon_\star$ is the dimensionless turbulent length scale indicating the expected length of the large-scale vortices, and $t_\star' = 1/\mathrm{Tt}_\star k_\star/\epsilon_\star$ is the dimensionless turbulent time scale describing the expected creation and break-up time of the vortices. For details on the general sampling procedure providing a fast and accurate sampling of the random fields we refer to \cite{huebsch:p:2013}. To even increase the efficiency of the procedure we use here a simplified underlying energy spectrum, see Appendix~\ref{appendix_turbRecon} for details on the modeling of $\mathbf{v}_{\star,loc}'$.
\begin{remark}[Lift forces]\label{rem:lift}
In industrial melt-blowing processes lift forces on a fiber are created through airflow vortices approaching the fiber and by vortex shedding at the back of the fiber. While the latter can be neglected since the fiber is meanly following the turbulent air stream, the first mechanism is included by the help of the following ansatz: the local turbulent instationary velocity fluctuations $\mathbf{v}_\star'$ are plugged into the air drag model (\ref{sec:model_eq:airdrag}), meaning local observations are mapped into a stationary far field consideration. This leads to aerodynamic forces on the fiber acting perpendicular to the $(\bar{\mathbf{v}}_\star-\mathbf{v})$-$\boldsymbol{\tau}$-plane.
\end{remark}

%
\setcounter{equation}{0} \setcounter{figure}{0} \setcounter{table}{0}
\section{Numerical Scheme}\label{sec:numerics}
System~\ref{system:Final} is a boundary value problem of a quasilinear system of partial differential equations of first order on a growing domain. It is discretized with finite volumes in space based on a central flux approximation with a Lax-Friedrich type stabilization and with the implicit Euler method in time. The growing fiber domain is realized by dynamic and static spatial cells according to the discretization concept in \cite{arne:p:2015}.

We reformulate System~\ref{system:Final} as
\begin{equation}\label{sec:numerics_eq:instationary}
\begin{aligned}
\mathbf{K}(\mathbf{y})\cdot\partial_t\mathbf{y} + \mathbf{L}(\mathbf{y})\cdot\partial_\zeta\mathbf{y} + \mathbf{l}(\mathbf{y}) = \mathbf{0}
\end{aligned}
\end{equation}
with the vector of unknowns $\mathbf{y} = (\mathbf{r},e,\vartheta,\varphi,\mathbf{v},T,\sigma,p)\in\mathbb{R}^{12}$ and consider it on the spatial domain $\mathcal{Q}(t) = (-t,0)$ for times $0 \leq t \leq t_{end}$. The introduction of the matrix $\mathbf{K}$ avoids a singularity for $\sin\vartheta = 0$. For $\sin\vartheta \neq 0$, $\mathbf{K}$ is invertible revealing the unique quasilinear form \eqref{sec:model_eq:quasilinearForm}.

\begin{figure}[!t]
\centering
\includegraphics[height=5.5cm]{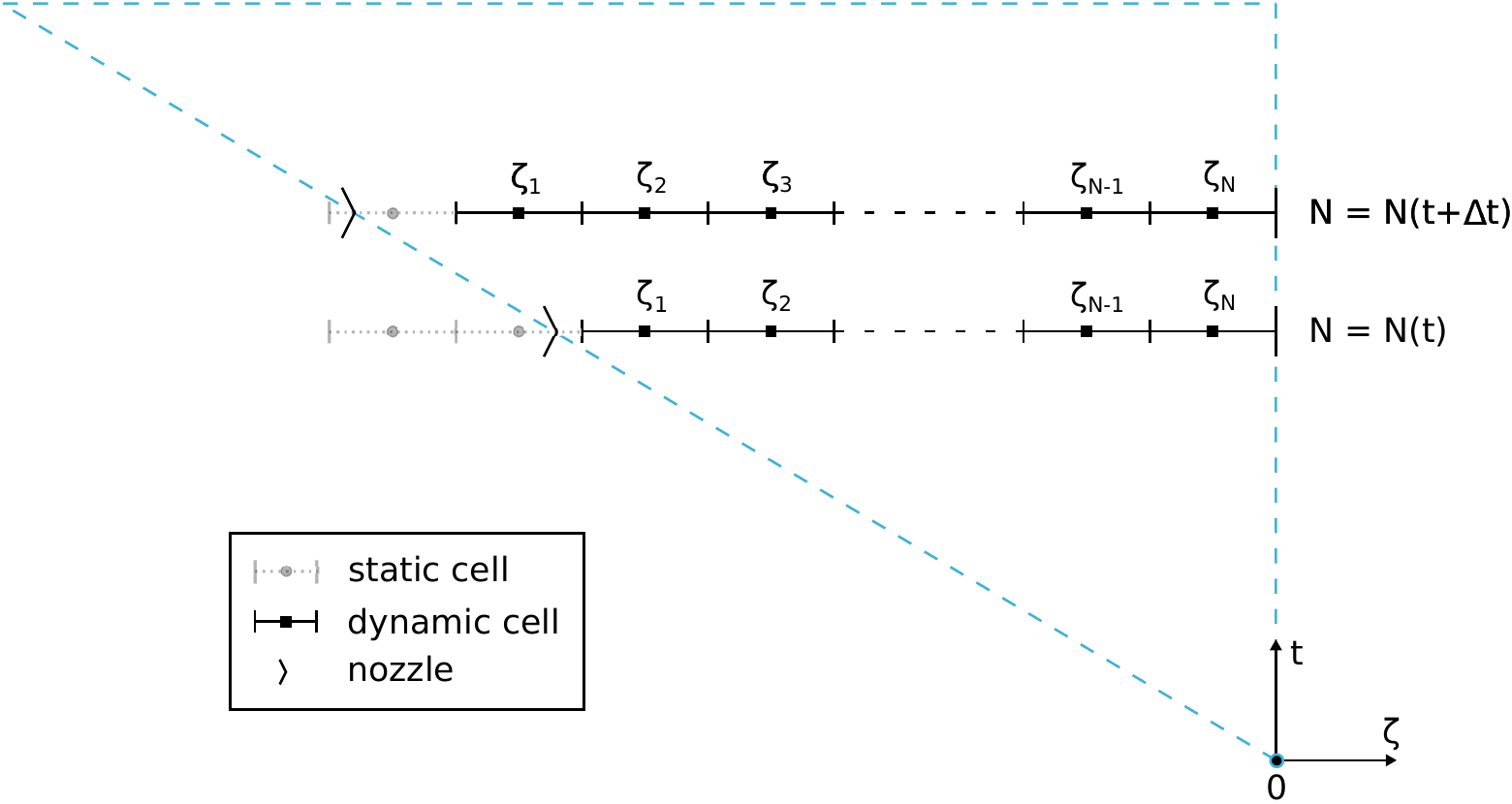}
\caption{Illustration of the spatial discretization of the growing jet domain $\mathcal{Q}(t)$ (marked by the blue dashed line) with $N(t)$ dynamic cells. Cells are treated as static cells until they have completly entered the flow domain.}\label{sec:numerics_fig:mesh}
\end{figure}

For the spatial discretization we employ a finite volume scheme. We introduce a constant cell size $\Delta \zeta$ and define the number of dynamic cells $N(t)$ depending on the fiber length $\zeta_L(t) = t$ at time $t$ as
\begin{align*}
N(t) = \bigg\lfloor\frac{\zeta_L(t)}{\Delta \zeta}\bigg\rfloor,
\end{align*}
where $\lfloor\cdot\rfloor$ denotes the floor function. Furthermore, we introduce the discretization points
\begin{align*}
\zeta_{(j+1)/2} = -\left( N(t) - \frac{j}{2}\right)\Delta \zeta, \qquad j=0,...,2N(t).
\end{align*}
The points $\zeta_i$, $i=1,...,N(t)$, represent the cell centers. The dynamic cell closest to the nozzle ($\zeta = -t$) is given by $[\zeta_{1/2}, \zeta_{3/2}]$, whereas $\zeta_{N+1/2} = 0$ is the fiber end, cf. Fig.~\ref{sec:numerics_fig:mesh}. The jet growth is realized by adding static cells at the nozzle. This means we add new cells, which are initialized by the boundary conditions at the nozzle (i.e., at the left side of the computational domain). The cells remain static until they have completely left the nozzle. When they have completely entered the flow domain they are called dynamic cells and are then taken into consideration for the computation. The introduction of static cells at the nozzle allows the suitable initialization of a jet with length $\zeta_L(t) < \Delta\zeta$ and a stable numerical treatment of the temporal evolution.

We define the cell averages $\mathbf{y}_i$, $i=1,...,N(t)$, of the unknown quantities as
\begin{align*}
\mathbf{y}_i(t) = \frac{1}{\Delta \zeta}\int\limits_{\zeta_{i-1/2}}^{\zeta_{i+1/2}} \mathbf{y}(\zeta,t)d\zeta,
\end{align*}
integrate the quasilinear system (\ref{sec:numerics_eq:instationary}) over the control cells $[\zeta_{i-1/2},\zeta_{i+1/2}]$, $i=1,...,N(t)$, in which we assume $\mathbf{X}(\mathbf{y})|_{[\zeta_{i-1/2},\zeta_{i+1/2}]} = \mathbf{X}(\mathbf{y}_i)$ for $\mathbf{X} = \mathbf{K}, \mathbf{L}, \mathbf{l}$ and adopt the idea of the Lax-Friedrichs scheme for the approximation of the numerical fluxes as done in \cite{fjordholm:p:2012}. The resulting system of ordinary differential equations for the cell averages $\mathbf{y}_i$ with respect to time has the form
\begin{equation}\label{sec:numerics_eq:spaceSemidis}
\mathbf{K}(\mathbf{y}_i)\cdot\frac{\mathrm{d}}{\mathrm{d}t}\mathbf{y}_i - \mathbf{K}(\mathbf{y}_i)\cdot\frac{1}{2\Delta t}(\mathbf{y}_{i+1} - 2\mathbf{y}_i + \mathbf{y}_{i-1})+ \mathbf{L}(\mathbf{y}_i)\cdot \frac{1}{2\Delta \zeta}\left(\mathbf{y}_{i+1} - \mathbf{y}_{i-1}\right) + \mathbf{l}(\mathbf{y}_i) = 0,
\end{equation}
where $\Delta t$ denotes the constant time-step size, that we will use in the temporal discretization. The incorporation of initial-boundary and boundary conditions in our numerical scheme is realized by ghostlayers. Following \cite{leveque:b:2002} quantities not being prescribed at a boundary are extrapolated on the corresponding ghostlayer, in particular we choose first order extrapolation.

For the solution of the system of ordinary differential equations (\ref{sec:numerics_eq:spaceSemidis}) we employ the stiffly accurate implicit Euler scheme with constant time-step size $\Delta t$
\begin{equation}\label{sec:num_eq:fullDiscr}
\begin{aligned}
\mathbf{K}(\mathbf{y}_i^{n+1})\cdot\left(2\mathbf{y}_i^{n+1} - \frac{1}{2}\mathbf{y}_{i+1}^{n+1} - \frac{1}{2}\mathbf{y}_{i-1}^{n+1} - \mathbf{y}_i^n\right) + \mathbf{L}(\mathbf{y}_i^{n+1})\cdot \frac{\Delta t}{2\Delta \zeta}\left(\mathbf{y}_{i+1}^{n+1} - \mathbf{y}_{i-1}^{n+1}\right) + \Delta t\mathbf{l}(\mathbf{y}_i^{n+1}) = 0,
\end{aligned}
\end{equation}
with $\mathbf{y}_i^n = \mathbf{y}_i(t^n)$ and $t^n = n\Delta t$ for $n = 0,...,M$, $t^M = t_{end}$.
The resulting nonlinear system of equations is solved by a Newton-method with Armijo step size control, where the Jacobian of the system matrix is prescribed analytically. The break-up criterion of the Newton algorithm is set to an absolute error tolerance $tol = 10^{-8}$ with respect to the maximum norm.
\begin{remark}[Artificial diffusion]
The semi-discrete system (\ref{sec:numerics_eq:spaceSemidis}) can be seen as a spatial discretization of
\begin{align*}
\mathbf{K}(\mathbf{y})\cdot\partial_t\mathbf{y} + \mathbf{L}(\mathbf{y})\cdot\partial_\zeta\mathbf{y} + \mathbf{l}(\mathbf{y}) = \mathbf{K}(\mathbf{y})\cdot\eta\partial_{\zeta\zeta}\mathbf{y}
\end{align*}
with $\eta = (\Delta \zeta)^2/(2\Delta t)$ by means of a central approximation of the flux terms. This means we add artificial diffusion of magnitude $\eta$ to our system as it is common for classical Lax-Friedrich schemes.
\end{remark}
\begin{remark}[Convergence of numerical scheme]
As it is well-known from hyperbolic literature (e.g. \cite{leveque:b:2002}), the numerical scheme (\ref{sec:num_eq:fullDiscr}) provides accuracy of order one with respect to the time and accuracy of order two with respect to the space discretization yielding a combined convergence of order one. In \cite{devuyst:p:2004, fjordholm:p:2012, munkejord:p:2009} a similar scheme has been investigated with respect to a stability concept for non-conservative hyperbolic partial differential equations.
\end{remark}
\begin{remark}[Spatial and temporal resolution]\label{sec:numerics_remark:resolution}
The temporal and spatial grid sizes have to be chosen in such a way that the turbulent scales of the underlying airflow are resolved properly. In particular, the turbulent length scale $l'_\star = \mathrm{Tu}_\star/\mathrm{Tt}_\star k_\star^{3/2}/\epsilon_\star$ and the turbulent time scale $t_\star' = 1/\mathrm{Tt}_\star k_\star/\epsilon_\star$ used in the turbulence reconstruction (\ref{sec:model_eq:turbReconstr}) have to be considered. Furthermore, the time that a vortex needs to pass a fixed material point of the fiber due to their relative velocity has to be taken into account for the temporal resolution. In total, the requirements for a successful simulation in terms of $\Delta\zeta$ and $\Delta t$ read
\begin{equation}\label{sec:numerics_eq:resolution}
\Delta\zeta \leq \frac{l'_\star}{e}, \qquad \Delta t \leq \min\left(t'_\star,\frac{l'_\star}{\lVert\mathbf{v_\star - \mathbf{v}}\rVert}\right).
\end{equation}
Appropriate grid sizes are estimated by computing the bounds for all given airflow data with assumptions on the maximal fiber velocity and elongation.
\end{remark}

%
\setcounter{equation}{0} \setcounter{figure}{0} \setcounter{table}{0}
\section{Industrial melt-blowing simulation}\label{sec:example}
In this section we investigate an industrial melt-blowing scenario that has been studied in \cite{huebsch:p:2013} by the help of a simplified ODE model for the fiber jet position, velocity and elongation. We employ our more sophisticated PDE fiber jet model (System~\ref{system:Final}), which additionally contains a viscoelastic material behavior and thermal effects describing the jet cooling and solidification. Before we present our simulation results, we specify the industrial setup and state the closing models for the dynamic viscosity as well as relaxation time and elastic modulus. In the scenario we face step size restrictions (cf.\ Remark~\ref{sec:numerics_remark:resolution}) that prevent the computability of the whole fiber from nozzle to conveyor belt. To handle this numerical problem we suggest and discuss an appropriate simulation strategy. 

\subsection{Setup and model closing}\label{sec:ex_subsec:setup}
\begin{table}[t]
\begin{minipage}[c]{\textwidth}
\begin{center}
\begin{small}
\begin{tabular}{| l l l l |}
\hline
\multicolumn{4}{|l|}{\textbf{Parameters}}\\
Description & Symbol & Value & Unit\\
\hline
device height & $H$ & $1.214\cdot10^{-1}$ & m \rule{0pt}{2.6ex}\\
nozzle diameter & $d_{in}$ & $4\cdot 10^{-4}$ & m\\
fiber speed at nozzle & $v_{in}$ &$1\cdot 10^{-2}$ & m/s\\
fiber temperature at nozzle & $T_{in}$ & $5.532\cdot10^2$ & K\\
heat transfer at nozzle & $\alpha_{in}$ & $ 1.595\cdot10^3$ & W/(m$^2$K)\\
polar angle at nozzle & $\vartheta_{in}$ & $\pi/2$ & --\\
azimuth angle at nozzle & $\varphi_{in}$ & $\pi$ & --\\
fiber density & $\rho$ & $7\cdot10^2$ & kg/m$^3$\\
fiber specific heat capacity &  $c_p$ & $2.1\cdot10^3$ & J/(kgK)\\
end time & $t_{end}$ & $2.20\cdot10^{-2}$ & s\\
air density at nozzle & $\rho_{\star,in}$ & $1.187$ & kg/m$^3$\\
air kinematic viscosity at nozzle & $\nu_{\star,in}$ & $1.8\cdot10^{-5}$ & m$^2$/s\\
air specific heat capacity at nozzle & $c_{p,\star,in}$ & $1.006\cdot10^3$ & J/(kgK)\\
air thermal conductivity at nozzle & $\lambda_{\star,in}$ & $2.42\cdot10^{-2}$ & W/(mK)\\
air turbulent kinetic energy at nozzle & $k_{\star,in}$ & $2.181\cdot10^2$ & m$^2$/s$^2$\\
air viscous dissipation rate at nozzle & $\epsilon_{\star,in}$ & $1.808\cdot10^7$ & m$^2$/s$^3$\\
\hline
\end{tabular}
\end{small}
\end{center}
\end{minipage}
\vfill
\vspace{0.5cm}
\begin{minipage}[c]{\textwidth}
\begin{center}
\begin{small}
\begin{tabular}{| l l l |}
\hline
\multicolumn{3}{|l|}{\textbf{Dimensionless numbers}}\\
Description & Symbol & Value\\
\hline
slenderness & $\varepsilon$ & $2.92\cdot 10^{-3}$ \rule{0pt}{2.6ex}\\
Reynolds & $\mathrm{Re}$ & $2.99\cdot 10^{-1}$\\
Deborah & $\mathrm{De}$ & $4.94\cdot 10^{-2}$\\
Froude & $\mathrm{Fr}$ & $9.16\cdot 10^{-3}$ \\
Stanton & $\mathrm{St}$ & $1.08\cdot 10^{-1}$\\
Mach & $\mathrm{Ma} $ & $1.22\cdot 10^{-1}$\\
air drag associated & $\mathrm{A}_\star$ & $5.81\cdot 10^{-1}$\\
mixed (air-fiber) Reynolds & $\mathrm{Re}_\star$ & $1.97\cdot 10^{-1}$\\
Nusselt & $\mathrm{Nu}_\star$ & $2.34\cdot 10^1$\\
Prandtl & $\mathrm{Pr}_\star$ & $8.89\cdot 10^{-1}$\\
turbulence degree & $\mathrm{Tu}_\star$ & $1.48\cdot 10^3$\\
turbulent time & $\mathrm{Tt}_\star$ & $1.01\cdot 10^6$\\
\hline
\end{tabular}
\end{small}
\end{center}
\end{minipage}\\~\\~\\
\caption{Overview over process and physical parameters in the industrial melt-blowing setup according to \cite{huebsch:p:2013} and the resulting dimensionless numbers.}\label{sec:ex_table:param}
\end{table}
\begin{figure}[!t]
\centering
\includegraphics[height=5cm]{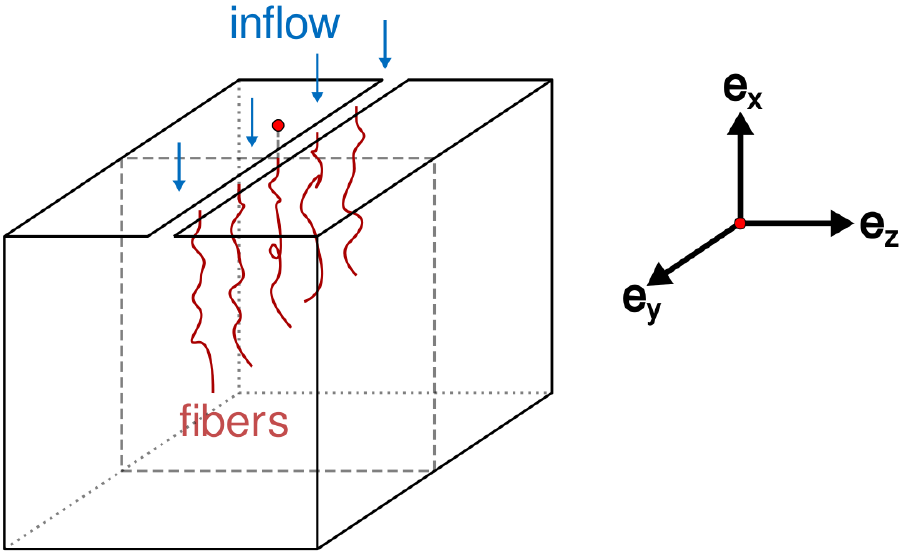}
\caption{Illustration of the considered industrial melt-blowing process. The two-dimensional cut ($\mathbf{e}_x$-$\mathbf{e}_z$-plane, marked by dashed line) represents the whole flow domain due to homogenity in $\mathbf{e}_y$-direction.}\label{sec:ex_fig:setup}
\end{figure}
\begin{figure}[!t]
\newlength\figureheight
\newlength\figurewidth
\centering
\begin{minipage}[c]{0.49\textwidth}
	\centering
	\includegraphics{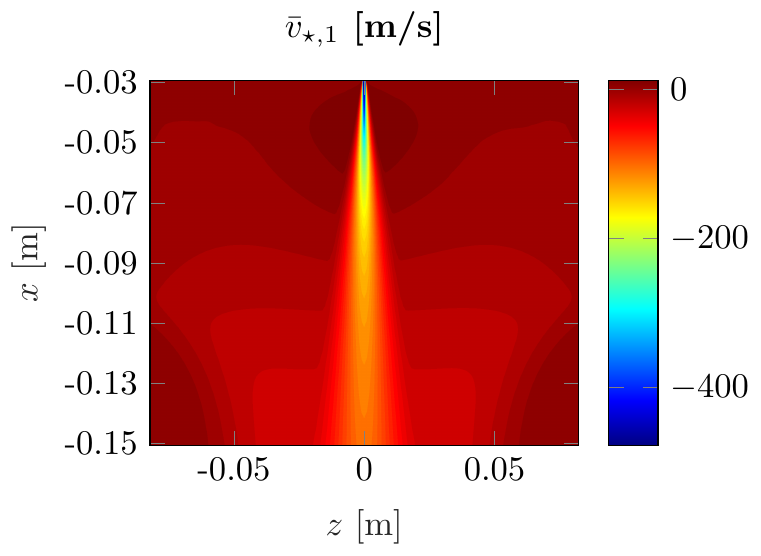}
\end{minipage}\hfill
\begin{minipage}[c]{0.49\textwidth}
	\centering
	\includegraphics{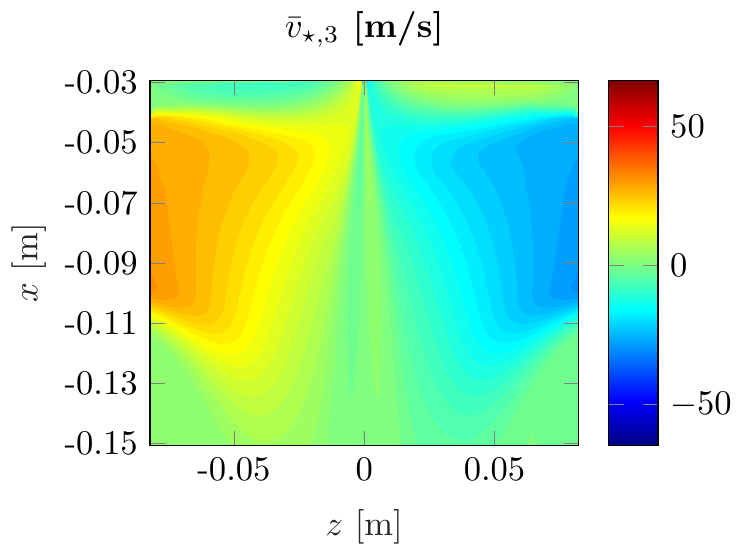}
\end{minipage}
\begin{minipage}[c]{0.49\textwidth}
	\centering
	\includegraphics{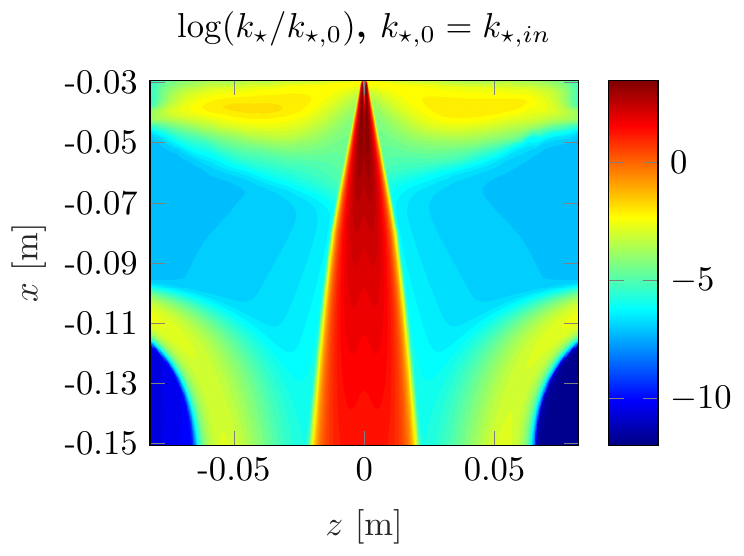}
\end{minipage}\hfill
\begin{minipage}[c]{0.49\textwidth}
	\centering
	\includegraphics{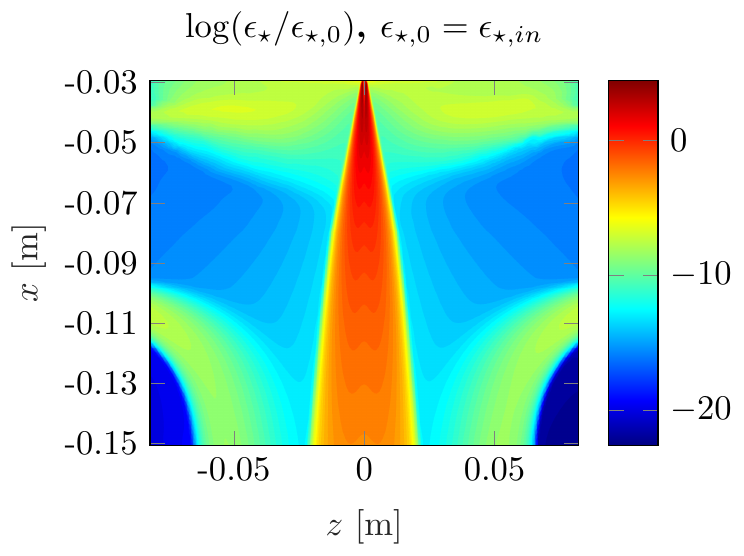}
\end{minipage}\vfill
\begin{minipage}[c]{0.49\textwidth}
	\centering
	\includegraphics{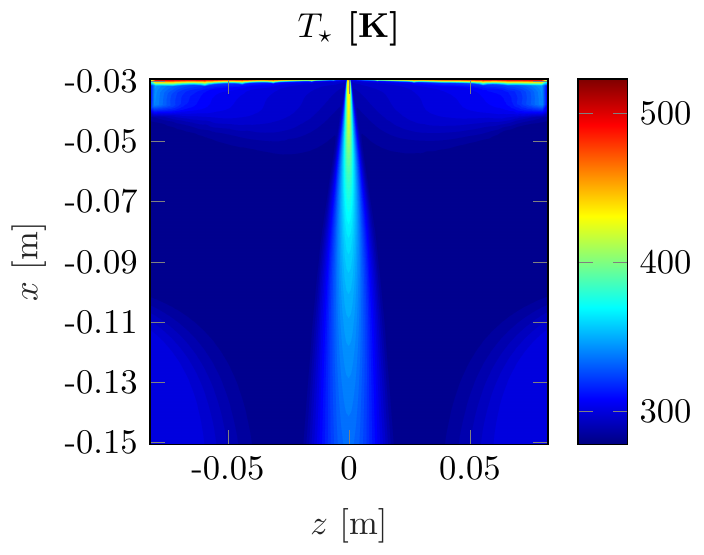}
\end{minipage}
\caption{Airflow simulation of the representative two-dimensional flow domain (cf. Fig.~\ref{sec:ex_fig:setup}). \textit{Top:} components of mean airflow velocity $\bar{\mathbf{v}}_\star$ in $\mathbf{e}_x$- and $\mathbf{e}_z$-direction (denoted by $\bar{v}_{\star,1}$ and $\bar{v}_{\star,3}$ respectively). \textit{Middle:} turbulent kinetic energy $k_\star$ and dissipation rate $\epsilon_\star$ in logarithmic scale. \textit{Bottom:} temperature $T_\star$.}\label{sec:ex_fig:k-eps-sim}
\end{figure}
In the melt-blowing setup a high-speed air stream is directed vertically downwards in direction of gravity and enters the domain of interest via thin slot dies. The spinning nozzles are located in between and extrude the polymeric fiber jets in the same direction, see Fig.~\ref{sec:ex_fig:setup}. We choose an outer orthonormal basis $\{\mathbf{e}_x,\mathbf{e}_y,\mathbf{e}_z\}$, where $\mathbf{e}_x$ points against the direction of gravity (i.e., $\mathbf{e}_x = -\mathbf{e}_g$) and $\mathbf{e}_y$ is aligned with the slot inlet. The mean quantities of the turbulent airflow are time-independent and homogeneous in $\mathbf{e}_y$-direction, such that a stationary $k_\star$-$\epsilon_\star$ simulation for a representative two-dimensional cut showing the $\mathbf{e}_x$-$\mathbf{e}_z$-plane is reasonable. The origin of the outer basis is aligned with the external given airflow data, such that the considered nozzle is at the position $\mathbf{r}_{in} = (-2.85\cdot 10^{-2},0,0)\,$m in the airflow field. We use the same $k_\star$-$\epsilon_\star$ simulation results as in \cite{huebsch:p:2013}, supplemented with an additional temperature profile as depicted in Fig.~\ref{sec:ex_fig:k-eps-sim}. The melt-blown fiber polymer is of polypropylene (PP) type with material parameters taken from \cite{huebsch:p:2013}. The process and physical parameters as well as the resulting dimensionless numbers are listed in Table~\ref{sec:ex_table:param}.

For the temperature-dependent dynamic viscosity of the PP-type fiber material we employ the Arrhenius law. The corresponding relation for the dimensionless viscosity $\mu$ depending on the temperature $T$ is given by
\begin{align*}
\mu(T) = \frac{1}{\mu_0}\mathcal{M}(TT_0),\qquad \mathcal{M}(T) = a_\mu\exp\left(\frac{b_\mu}{T-c_\mu}\right).
\end{align*}
The polymer-specific constants coming from measurements are $a_\mu = 0.1352$ Pas, $b_\mu = 852.323$ K, $c_\mu = 273.15$ K.
We choose the following heuristic model for the relaxation time
\begin{align*}
\theta(T) = \frac{1}{\theta_0} \mathcal{T}(TT_0),\qquad  \mathcal{T}(T) = 3\frac{\mathcal{M}(T) + b_\theta}{a_\theta}
\end{align*}
with $a_\theta=10^9$ Pa and $b_\theta= 2\cdot 10^8$ Pas showing a meaningful limit behavior: for $T\rightarrow\infty$ the dimensional relaxation time is of order $\mathcal{T}\sim\mathcal{O}(10^{-1}$ s), which is typical for melt-blown polymers, see, e.g., \cite{yarin:p:2010b}. Furthermore, employing the relation $\mu/\theta = E/3$ the resulting dimensionless elastic modulus $E$ reads
\begin{align*}
E(T) = \frac{\theta_0}{\mu_0}\mathcal{E}(T_0T),\qquad \mathcal{E}(T) = 3\frac{\mathcal{M}(T)}{\mathcal{T}(T)}.
\end{align*}
For $T\rightarrow c_\mu = 273.15$ K the dimensional elastic modulus $\mathcal{E}$ approaches $\mathcal{E} = a_\theta = 10^9$ Pa -- a typical value for hardened polymer, see, e.g., \cite{barnes:b:2000}.

\subsection{Simulation strategy}\label{sec:ex_subsec:practicalTreatment}
Expecting a maximal fiber elongation $e = 10^{6}$ and a maximal dimensionless relative velocity between fiber and airflow $\lVert \mathbf{v}_\star-\mathbf{v} \rVert \leq \lVert \mathbf{v}_\star \rVert = 4.78\cdot 10^4$ (in dimensionless form) in the industrial melt-blowing, the  step size restriction for the spatial and temporal fiber discretization (\ref{sec:numerics_eq:resolution}) gives
\begin{align}\label{sec:ex_eq:resolution}
\Delta \zeta \leq 5.77\cdot 10^{-10},\qquad \Delta t \leq 1.21\cdot 10^{-6}
\end{align}
(cf. Fig.~\ref{sec:ex_fig:k-eps-sim}). Such a resolution implies computationally impractical runtimes. However, to make a simulation for the setup feasible, we suggest the following strategy that is motivated from observations of the process.

In the region close to the nozzle the high-speed air stream pulls the slowly extruded fiber jet rapidly down without any lateral bending. The hot temperatures prevent fiber cool-down and solidification. Thus, the magnitude of the Deborah number $\mathrm{De}$ at the nozzle (cf. Tab.~\ref{sec:ex_table:param}) allows the consideration of the viscous limit case $\mathrm{De}\rightarrow 0$ (see Remark~\ref{sec:model_remark:viscousLimit}). Moreover, the fiber jet behavior is mainly determined by the mean airflow, turbulent effects play a negligible role. Hence, we assume that in the nozzle region (i.e., deterministic region) the polymer jet can be described by a steady uni-axial viscous fiber model with deterministic aerodynamic force and heat transfer. This model follows from System~\ref{system:Final} by a re-parameterization into Euler (spatial) description, transition to steady-state, and the limit $\mathrm{De}\rightarrow 0$. The resulting boundary value problem of ordinary differential equations is solved by a continuation-collocation method, which we successfully employed in studies on glass wool manufacturing \cite{arne:p:2011}, electrospinning \cite{arne:p:2018} and dry spinning \cite{wieland:p:2018b}. Further details on the model and its numerical treatment are given in Appendix~\ref{appendix_statModel}. Note that the use of the viscous fiber model is not only physically reasonable, but it also simplifies crucially the numerical treatment. Concerning the viscoelastic fiber model, the rapid changes of the fiber quantities in the nozzle region caused by the immediate pull down of the fiber yields multiple changes in the structure of the quasilinear system matrix in view of its eigenvalues and its resulting classification. This means that the runs of the characteristics change their direction several times. In the steady uni-axial model this leads to singular system matrices and closing problems with appropriate boundary conditions making the numerical treatment extremely complicated. This issue has been addressed by \cite{lorenz:p:2014} in the context of existence regimes for solutions of an uni-axial UCM fiber model under gravitational forces. We circumvent these problems when using the viscous fiber model where no mathematical regime changes take place. 

In the region away from the nozzle the turbulent aerodynamic fluctuations crucially affect the fiber behavior (i.e., stochastic region). By means of the uni-axial steady fiber solution (from the nozzle region) we identify a coupling point, from where on the further fiber behavior downwards to the bottom is described by the instationary viscoelastic fiber model (System~\ref{system:Final}) accounting for turbulent effects. The simulation with the numerical scheme from Sec.~\ref{sec:numerics} becomes here feasible since the expected elongation and relative velocities in this domain are much smaller and hence the spatial and temporal step size restrictions weaken compared to (\ref{sec:ex_eq:resolution}). 

The coupling between the stationary and instationary fiber simulations is done in the following way: Let the fiber domain in the Eulerian parametrization $\Omega(t) = \Omega_d\cup\Omega_s(t)$ be divided into the time-independent deterministic part $\Omega_d$, where the fiber is uni-axially stretched, and the time-dependent stochastic part $\Omega_s(t)$, where the fiber is strongly affected by the turbulent fluctuations. Consider $C = \Omega_d\cap\Omega_s(t)$ to be the time-independent coupling point between the deterministic and the stochastic domain. 
First, we perform the simulation of the steady viscous fiber model (System~\ref{system:Final_viscous} in Appendix~\ref{appendix_statModel}) for the whole fiber domain, i.e., $\Omega_d = \Omega$, yielding solutions for the scalar fiber speed $u$, temperature $T$, stress $\sigma$ and pressure $p$. Second, we determine the coupling point $C$ by the ratio of the relative velocity between fiber and airflow $v_{rel} = \lVert \mathbf{v}_\star - u\boldsymbol{\tau} \rVert$ and the turbulent velocity scale $k_\star^{1/2}$, in particular
\begin{align*}
C=\min\bigg\{s\in\Omega~\bigg\lvert\left(\frac{v_{rel}v_0}{(k_{\star}k_{\star,0})^{1/2}}\right)(s) \leq 10\bigg\}.
\end{align*}
So the coupling point is the nearest point to the nozzle, where the ratio of the relative velocity and the turbulent velocity scale is below one order of magnitude. At $C$ the quantities of the stationary solution are denoted by $u_C$, $T_C$, $\sigma_C$, $p_C$. Third, for the subsequent solving of the instationary viscoelastic fiber model (System~\ref{system:Final}) on $\Omega_s(t)$ (reformulated in Lagrangian coordinates) we adjust the typical values and adapt the initial conditions. We particularly set the reference values used for the non-dimensionalization (see Sec.~\ref{sec:model_general}) to be
\begin{align*}
r_0 = (1-C)H,\qquad d_0 =\frac{\sqrt{\pi}}{2} \sqrt{\frac{v_{in}}{v_C}} d_{in},\qquad
v_0 = u_C, \qquad  T_0 = T_C, \end{align*}
 then the dimensionless numbers change accordingly. The altered initial conditions read
\begin{align*}
\sigma_{in} = \sigma_C, \qquad p_{in} = p_C.
\end{align*}
These modifications can be interpreted as putting a fictive nozzle with adjusted extrusion conditions at the spatial position of the coupling point $C$. The diameter of the fictive nozzle reflects the pre-elongations of the extruded fiber by the factor $v_C/v_{in}$ compared to $d_{in}$ of the original nozzle.

In the setup the crucial stretching of the fiber takes place in the upper part of the device and ends when the fiber is nearly solidificated. Since we are interested in the maximal achieved fiber elongations as well as in the corresponding fiber diameter distribution, but not in the lay-down process, it is sufficient to cutoff the fiber before it reaches the bottom of the device. We choose to cutoff the fiber, when it reaches the height corresponding to $x = -9.45\cdot10^{-2}$ m. Below this point the airflow temperature satisfies $T_\star < 353.15$ K (see Fig.\ \ref{sec:ex_fig:k-eps-sim}). We expect the fiber dynamic viscosities to be of magnitude $\mu\sim\mathcal{O}(10^3$ Pas$)$, implying that no noticeable further fiber elongations take place.
\begin{figure}[!t]
\centering
\begin{minipage}[c]{\textwidth}
	\centering
	\includegraphics{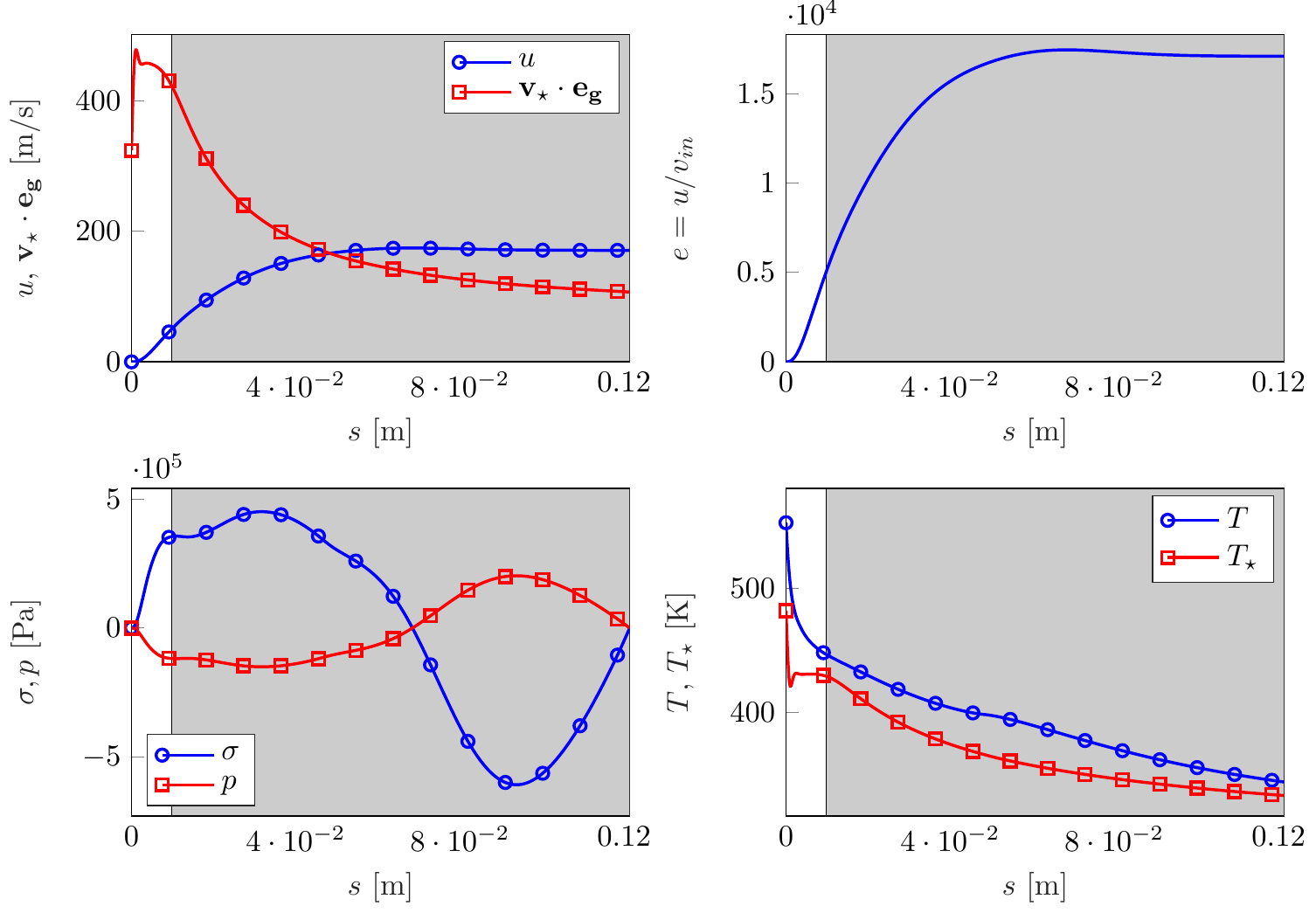}
\end{minipage}
\caption{Scalar speed $u$, elongation $e$, stress $\sigma$, pressure $p$ and temperature $T$ of the steady viscous uni-axial fiber model in Eulerian coordinates \textit{(top left to bottom right)}. The stochastic region, where the instationary viscoelastic fiber model (System~\ref{system:Final}) is employed, is shaded in gray.}\label{sec:ex_fig:statSol}
\end{figure}

\subsection{Results}
In the following we present the numerical results for the industrial spinning setup described in Sec.~\ref{sec:ex_subsec:setup}. While all computations for a single fiber realization have been done on an Intel Core i7-6700 CPU (4 cores, 8 threads) and 16 GBytes of RAM, the Monte Carlo simulation has been performed on a MPI cluster (dual Intel Xeon E5-2670, 16 CPU cores per node, 64 GB RAM) with one CPU core for each fiber realization. For all computations the MATLAB version R2016b has been used.

Figure~\ref{sec:ex_fig:statSol} shows the results for scalar speed $u$, stress $\sigma$, pressure $p$ and temperature $T$ as well as the induced elongation $e$ of the steady viscous uni-axial fiber model in an Eulerian parameterization that spans the whole domain $\Omega$. The maximal fiber speed is $u=1.74\cdot10^2$ m/s and the corresponding maximal fiber elongation is $e=1.74\cdot10^4$. This indicates that a stationary fiber simulation is not physically reasonable for the whole domain $\Omega$, since much higher fiber elongations for the melt-blowing setup are expected. Nevertheless, the steady viscous solution serves as adequate approximation of the fiber behavior in the nozzle region as described in Sec.~\ref{sec:ex_subsec:practicalTreatment}. For the further instationary viscoelastic simulation we determine the spatial position of the coupling point $C$ and put a fictive nozzle at $\mathbf{r}_{in} = (-3.83\cdot10^{-2},0,0)\,$m. The corresponding fiber quantities at this fictive nozzle are
\begin{align*}
u_C = 50.62\text{ m/s}, \qquad \sigma_C = 69.98\text{ Pa},\qquad p_C = -23.33\text{ Pa},\qquad T_C = 446.6\text{ K},
\end{align*}
and the dimensionless numbers change accordingly, see Tab.~\ref{sec:ex_table:dimNumbers_adjustedNozzle}.
\begin{table}[t]
\begin{minipage}[c]{\textwidth}
\begin{center}
\begin{small}
\begin{tabular}{| l l l |}
\hline
\multicolumn{3}{|l|}{\textbf{Dimensionless numbers}}\\
Description & Symbol & Value\\
\hline
slenderness & $\varepsilon$ & $4.47\cdot10^{-5}$ \rule{0pt}{2.6ex}\\\
Reynolds & $\mathrm{Re}$ & $2.15\cdot10^2$ \\
Deborah & $\mathrm{De}$ & $2.72\cdot10^2$\\
Froude & $\mathrm{Fr}$ & $4.84\cdot10^1$ \\
Stanton & $\mathrm{St}$ & $1.75\cdot10^{-4}$\\
Mach & $\mathrm{Ma}$ & $2.42\cdot10^{2}$\\
air drag associated & $\mathrm{A}_\star$ & $2.66\cdot10^1$\\
mixed (air-fiber) Reynolds & $\mathrm{Re}_\star$ & $1.40\cdot 10^1$\\
Nusselt & $\mathrm{Nu}_\star$ & $2.68$\\
Prandtl & $\mathrm{Pr}_\star$ & $6.23\cdot 10^{-1}$\\
turbulence degree & $\mathrm{Tu}_\star$ & $7.34\cdot10^{-1}$\\
turbulent time & $\mathrm{Tt}_\star$ & $1.14\cdot10^2$\\
\hline
\end{tabular}
\end{small}
\end{center}
\end{minipage}\\~\\~\\
\caption{Dimensionless numbers characterizing the fiber behavior in the stochastic region.}\label{sec:ex_table:dimNumbers_adjustedNozzle}
\end{table}
\begin{figure}[!t]
\centering
\begin{minipage}[c]{0.33\textwidth}
	\centering
	\includegraphics{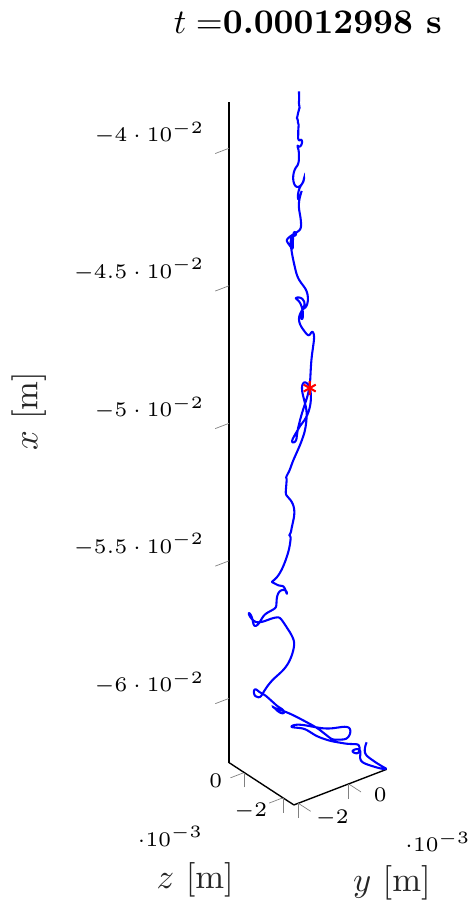}
\end{minipage}\hfill
\begin{minipage}[c]{0.33\textwidth}
	\centering
	\includegraphics{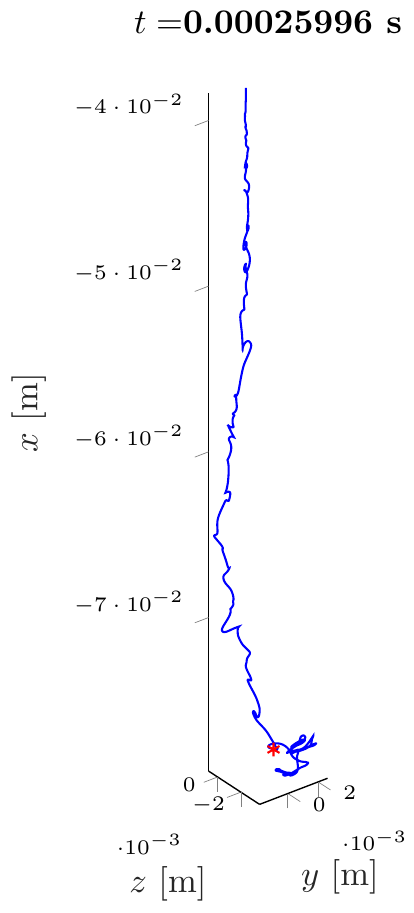}
\end{minipage}\hfill
\begin{minipage}[c]{0.33\textwidth}
	\centering
	\includegraphics{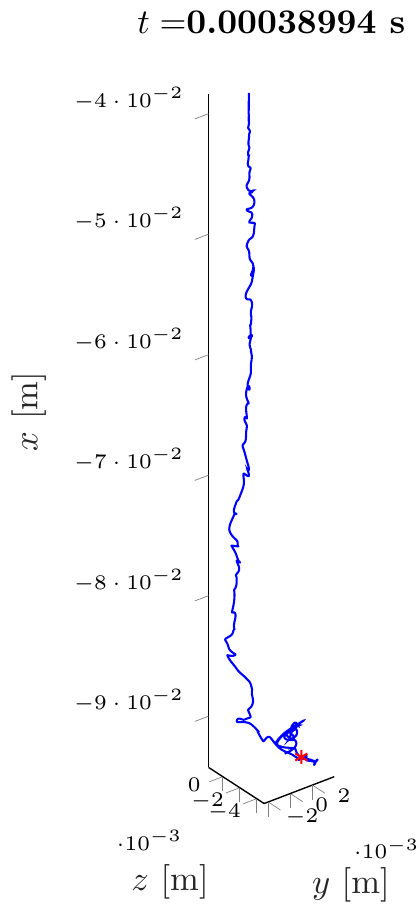}
\end{minipage}\hfill
\caption{Snapshots of one representative fiber curve $\mathbf{r}$ before reaching the cutoff height ($x = -9.45\cdot10^{-2}$ m) at times $t \in\{ 1.30\cdot10^{-4}$ s, $2.60\cdot10^{-4}$ s, $3.90\cdot10^{-4}$ s$\}$. We track the material point $\zeta_{N-3269}$ (marked with a red star) and present the temporal evolution of all fiber quantities at that point in Fig.~\ref{sec:ex_fig:sol_instat}.}\label{sec:ex_fig:curves}
\end{figure}
\begin{figure}[t]
\begin{minipage}[c]{\textwidth}
	\centering
	\includegraphics{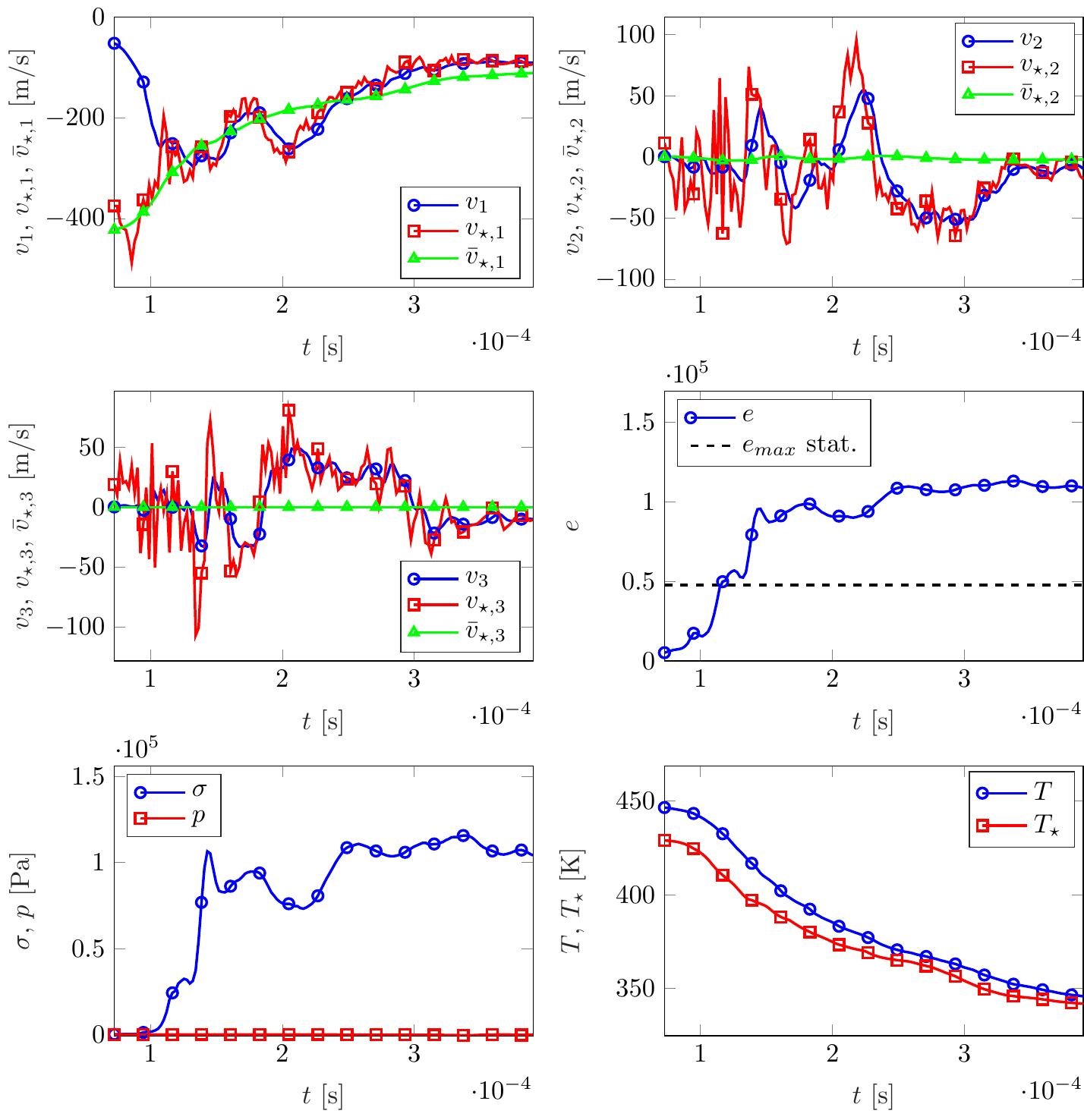}
\end{minipage}
\caption{Solution plots for the material point $\zeta_{N-3269}$ (cf. Fig.~\ref{sec:ex_fig:curves}) that enters the flow domain at time step $3270$ ($t = 7.20\cdot10^{-5}$ s). \textit{Top} and \textit{middle left:} fiber velocities $v_i$ (\textit{blue}) as well as airflow velocities $v_{\star,i}$, $\bar{v}_{\star,i}$ with (\textit{red}) and without (\textit{green}) turbulent fluctuations respectively, $i \in\{ 1,2,3\}$. \textit{Middle right:} elongation $e$ as rate of fiber stretching compared to the original nozzle and $e_{max}$ indicating the maximal achievable elongation in stationary simulations. \textit{Bottom:} pressure $p$, stress $\sigma$ as well as fiber temperature $T$ and air temperature $T_\star$.}\label{sec:ex_fig:sol_instat}
\end{figure}
\begin{figure}[!t]
\centering
\begin{minipage}[c]{0.49\textwidth}
	\centering
	\includegraphics{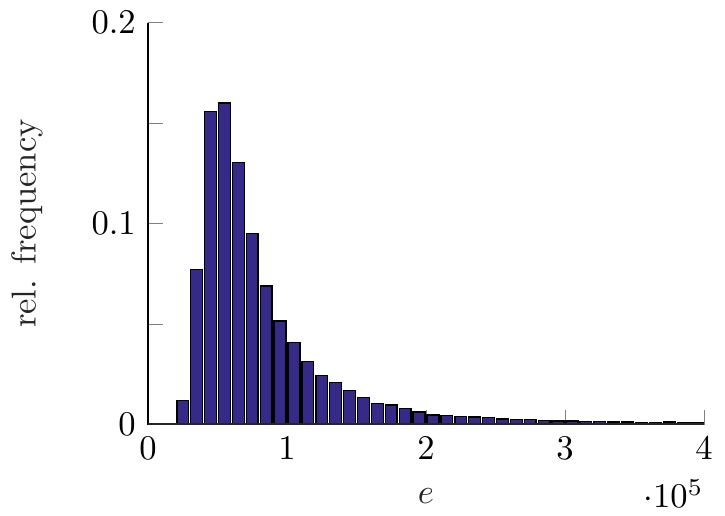}
\end{minipage}\hfill
\begin{minipage}[c]{0.49\textwidth}
	\centering
	\includegraphics{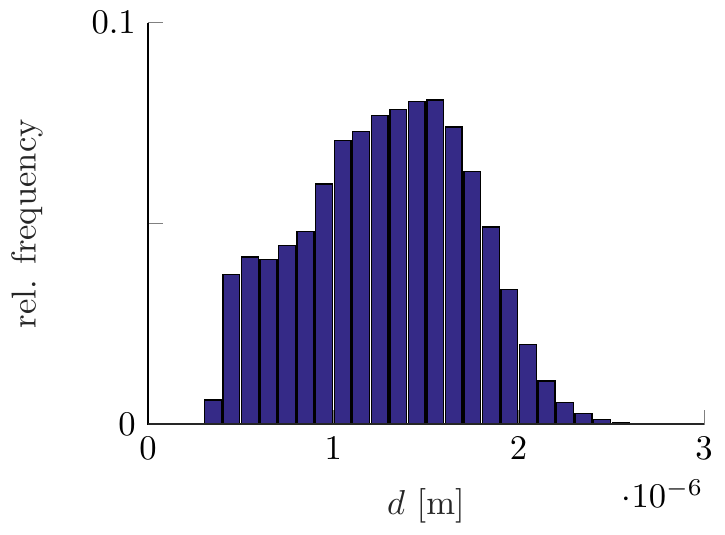}
\end{minipage}
\caption{\textit{Left:} Fiber elongation distribution at the cutoff point for a Monte Carlo simulation based on $67$ realizations. 
\textit{Right:} Resulting fiber diameter distribution at the cutoff point in the sense of an Eulerian fiber parameterization.
}\label{sec:ex_fig:diameter}
\end{figure}

Considering the stochastic region, the numerical step size restriction (\ref{sec:numerics_eq:resolution}) for the fiber discretization weakens compared to (\ref{sec:ex_eq:resolution})
\begin{align*}
\Delta\zeta \leq 3.31\cdot 10^{-5}, \qquad \Delta t \leq 7.99\cdot 10^{-4},
\end{align*}
we choose $\Delta\zeta = \Delta t = 10^{-5}$ for our computation. As expected the turbulent fluctuations of the airflow cause a swirling of the fiber jet such that the fiber curve leaves the $\mathbf{e}_x$-axis shortly away from the fictive nozzle. Figure~\ref{sec:ex_fig:curves} shows temporal snapshots of the curve for one representative fiber before its cutoff (at $x = -9.45\cdot10^{-2}$ m). The fluctuations move the fiber jet not only downwards but also upwards such that the fiber curve creates loops. In these loops high aerodynamic forces act on the fiber due to high relative velocity gradients causing the fiber to elongate. Figure~\ref{sec:ex_fig:sol_instat} shows exemplary the temporal evolution of the fiber quantities for one material point. Obviously the material point experiences high elongations: directly after entering the flow domain high relative velocities in $\mathbf{e}_x$-direction between the fiber velocity $v_1$ and the deterministic airflow velocity $\bar{v}_{\star,1}$ cause a fiber stretching. After the fiber velocity $v_1$ reaches the corresponding deterministic airflow velocity $\bar{v}_{\star,1}$ the fiber experiences a further stretching due to the velocity fluctuations, in which the mean stretching takes place in regions where high lateral air velocities $v_{\star,2}$, $v_{\star,3}$ create swirls. The final elongation at this material point is of magnitude $e\sim\mathcal{O}(10^5)$ and therewith clearly exceeds the theoretically possible deterministic expectations. In particular, the computed elongation in a stationary simulation is obviously restricted by the velocity of the air stream, i.e., $e_{max} = u/v_{in} < \lVert \mathbf{v}_\star \rVert_\infty/v_{in} = 4.78\cdot10^4$. Furthermore, the stationary uni-axial viscous simulation only achieves $e = 1.74\cdot 10^4$ (cf. Fig.~\ref{sec:ex_fig:statSol}). In the region of high fiber stretching the material point experiences high stresses $\sigma$ that partly dissipate due to the elastic material behavior before the fiber completely solidificates. The pressure $p$ is orders of magnitude smaller compared to the stress $\sigma$ and could therefore be neglected in the simulation as already pointed out in Remark~\ref{sec:model_remark:p}. The fiber temperature $T$ approaches the air temperature $T_\star$ leading to a cool-down and induced solidification of the jet.
 
When the fiber reach the cutoff height $x = -9.45\cdot10^{-2}$~m at time $t = 3.92\cdot10^{-4}$~s, we cutoff the fiber end, track the fiber elongations $e$ as well as the corresponding fiber diameters $d$ and document the occurring relative frequencies until the end time $t_{end} = 2.20\cdot10^{-2}$~s is reached, see Fig.~\ref{sec:ex_fig:diameter}. To achieve comparability with experiments, we weight the relative frequencies of the fiber diameters with the associated fiber elongations $e$ leading to a diameter distribution in the sense of an Eulerian (spatial) parameterization of the fibers. The resulting elongation and fiber diameter distributions are computed by the help of a Monte Carlo simulation with $67$ samples. We observe a mean elongation $e = 9.47\cdot10^4$ again exceeding the deterministic expectations. The mean fiber diameter is $d = 1.28\cdot10^{-6}$~m. This is a typical value for fibers produced in industrial melt-blowing setups, see for example \cite{ellison:p:2007}. So our instationary viscoelastic fiber model using an adjusted nozzle as well as employing fluctuation reconstruction of the underlying turbulence effects from an airflow simulation predicts quantitatively well the fiber jet thinning observed in experiments, which would not be possible with a pure steady deterministic simulation neglecting the turbulent aerodynamics velocity fluctuations.

Summing up, our proposed procedure makes the simulation of industrial melt-blowing processes with inclusion of turbulent and viscoelastic effects as well as temperature dependencies feasible. Including turbulent effects acting on the fiber by the help of reconstructing the turbulent structure of the outer air stream yields a jet thinning exceeding the deterministic expectations and produces final fiber diameters of realistic order of magnitude. So our presented modeling and solution framework provides the basis for further parameter studies and the optimization of melt-blowing processes. The computation time for the presented setup is around $96.4$ hours. A combined experimental and numerical study is left to future research.

\section{Conclusion}
In this paper we presented a model and simulation framework that allowed the numerical investigation of the physical mechanism being responsible for the strong fiber thinning in industrial melt-blowing processes. Considering an asymptotic instationary viscoelastic UCM fiber jet model driven by turbulent aerodynamic forces, the random field sampling strategy of \cite{huebsch:p:2013} provides an efficient numerical procedure for the realization of the turbulent air flow fluctuctuations. The computed fiber diameters are much lower than those obtained from previous stationary simulations regarding a pure deterministic aerodynamic force on the fiber. Our simulation results clearly stress the significance of the turbulent effects as key player for the production of fibers of micro- and nanoscale.  Further parameter studies and an optimization of the industrial process setup will provide the opportunity of simulating fibers with elongations of order $e\sim\mathcal{O}(10^6)$ compared to the nozzle diameter. In view of more quantitative predictions of the resulting nonwovens a combined experimental and numerical study with experimentally measured temperature-dependencies of polymer properties (e.g., relaxation time) is aimed at in future.

\section*{Acknowledgments}
This work has been supported by German DFG, project 251706852, MA 4526/2-1, WE 2003/4-1.
\appendix
\renewcommand*{\thesection}{\Alph{section}}
\renewcommand\thefigure{\thesection.\arabic{figure}}
\renewcommand\theequation{\thesection.\arabic{equation}}
\setcounter{section}{0}
\setcounter{figure}{0}
\setcounter{equation}{0}
\section{Exchange models between airflow and fiber}
The exchange models between airflow and fiber (Sec.~\ref{subsubsec:drag}) depend on the (in-)flow situation prescribed by the fiber orientation (normalized tangent) $\mathbf{t}$ and the relative velocity between airflow and fiber $\mathbf{w}$. They go back to studies of a stationary perpendicular laminar flow around a cylinder and have been extended to cover arbitrary angle of attacks and velocity regimes.
\subsection{Air resistance coefficients} \label{appendix_AirDrag}
The dimensionless air drag function $\mathbf{F}:\mathrm{SO}(3)\times \mathbb{R}^3\rightarrow \mathbb{R}^3$, $\mathbf{F}(\mathbf{t},\mathbf{w}) = r_n(w_n)\mathbf{w_n} + r_t(w_n)\mathbf{w_t}$ can be expressed in terms of the normal $\mathbf{w_n}=\mathbf{w}-\mathbf{w_t}$, $w_n = \lVert \mathbf{w_n} \rVert$ and tangential $\mathbf{w_t} = (\mathbf{w}\cdot\mathbf{t})\mathbf{t}$ relative velocity components. For the air resistance coefficients $r_n$, $r_t$ we use the following model taken from \cite{marheineke:p:2009b}
\begin{align*}
r_n(w_n) &= \begin{cases}
\sum_{j=0}^3 q_{n,j}w_n^j,\quad & w_n < w_0,\\
\frac{4\pi}{S(w_n)}\big(1-\frac{S^2(w_n)-S(w_n)/2+5/16}{32S(w_n)}w_n^2\big),\qquad \quad &w_0 \leq w_n < w_1,\\
w_n\exp\big(\sum_{j=0}^3 p_{n,j}\log^j(w_n)\big), &w_1 \leq w_n \leq w_2,\\
2\sqrt{w_n}+0.5w_n, &w_2 < w_n,
\end{cases}\\
r_t(w_n) &= \begin{cases}
\sum_{j=0}^3 q_{t,j}w_n^j,\quad &w_n < w_0,\\
\frac{4\pi}{(2S(w_n)-1)}\big(1-\frac{2S^2(w_n)-2S(w_n)+1}{16(2S(w_n)-1)}w_n^2\big),\qquad \, &w_0 \leq w_n < w_1,\\
w_n\exp\big(\sum_{j=0}^3 p_{t,j}\log^j(w_n)\big), &w_1 \leq w_n \leq w_2,\\
2\sqrt{w_n}, &w_2 < w_n,
\end{cases}
\end{align*}
with $S(w_n) = 2.0022 - \log(w_n)$. It is composed of asymptotic Oseen theory, Taylor heuristic and simulations where the transition points $w_1 = 0.1$ and $w_2 = 100$ are estimated from a least-square approximation of experimental and numerical data. For tangential incident flow situations ($w_n \rightarrow 0$) a regularization employs the Stokes theory, yielding the Stokes limits
\begin{align*}
r_n^S = \frac{4\pi}{\log(4\epsilon^{-1})}-\frac{\pi}{\log^2(4\epsilon^{-1})},\qquad r_t^S = \frac{2\pi}{\log(4\epsilon^{-1})}+\frac{\pi}{2\log^2(4\epsilon^{-1})},
\end{align*}
as well as the transition point $w_0 =2\exp\big(2.0022-{4\pi}/{r_n^S}\big)$ with the regularization parameter $\epsilon = 3.5\cdot 10^{-2}$. The parameters $p_{k,j}$ and $q_{k,j}$ ($k \in \{n,t\}$, $j\in\{0,1,2,3\}$) ensure smoothness,
\begin{alignat*}{6}
p_{n,0} &= 1.6911, \qquad &p_{n,1} &= -6.7222\cdot 10^{-1},\qquad
&p_{n,2} &= 3.3287\cdot 10^{-2},\qquad &p_{n,3}&=3.5015\cdot 10^{-3},\\
p_{t,0} &= 1.1552,\qquad &p_{t,1} &= -6.8479\cdot 10^{-1},\qquad
&p_{t,2} &= 1.4884\cdot 10^{-2},\qquad &p_{t,3}&=7.4966\cdot 10^{-4},
\end{alignat*}
\begin{align*}
q_{k,0} &= r_k^S,\qquad q_{k,1} = 0,\qquad q_{k,2} = \frac{3r_k(w_0)-w_0r_k'(w_0)-3r_k^S}{w_0^2},\\
 q_{k,3} &= \frac{-2r_k(w_0)+w_0r_k'(w_0)+2r_k^S}{w_0^3}.
\end{align*}
\setcounter{figure}{0}
\setcounter{equation}{0}
\subsection{Heat transfer coefficient}\label{appendix_Nusselt}
For the heat transfer coefficient $\alpha$ we model the Nusselt number associated function $\mathcal{N}:\mathbb{R}^3\rightarrow\mathbb{R}$ as
\begin{align*}
\mathcal{N}(w_t,w,p) = \left(1-\frac{1}{2}\,{h}(w_t, w)\right)
\begin{cases}
{n_1}(w,p),\quad & wp\geq \delta,\\
{n_2}(w,p),\quad & wp < \delta,
 \end{cases}
\end{align*}
depending on the tangential and absolute relative velocity ($w_t=\|\mathbf{w_t}\|$, $w=\|\mathbf{w}\|$) and the Prandtl number, with $\delta = 7.3\cdot 10^{-5}$. The model goes back to \cite{sucker:p:1976} where originally a stationary perpendicular laminar flow situation ($w_t=0$) around a cylinder for $wp \geq \delta$ was studied and has been extended to ensure a smooth transition to the limit value $\mathcal{N}\rightarrow\mathcal{N}_0=0.1$ for vanishing $wp\rightarrow 0$,
\begin{align*}
{n_1}(w,p) = a_\mathcal{N}(wp)^{0.1} + f(p)\frac{(wp)^{0.7}}{1+b_\mathcal{N}(wp)^{0.2}},\qquad 
{n_2}(w,p) = m_1(p)(wp)^3 + m_2(p)(wp)^2 + \mathcal{N}_0,
\end{align*}
with the coefficients
\begin{align*}
m_1(p) = c_\mathcal{N} + d_\mathcal{N} f(p),\qquad m_2(p) = e_\mathcal{N} + g_\mathcal{N} f(p),\qquad f(p) = \frac{k_\mathcal{N}}{(1 + (l_\mathcal{N}p^{1/6})^{2.5})^{0.4}}
\end{align*}
and the constant parameters
\begin{align*}
a_\mathcal{N} &= 0.462, \qquad &b_\mathcal{N} &= 2.79, \qquad &c_\mathcal{N} &= -3.5636\cdot 10^{11}, \qquad &d_\mathcal{N} &= - 3.1380\cdot 10^9, \\
e_\mathcal{N} &= 4.0694\cdot 10^7, \qquad &g_\mathcal{N} &= 4.0694\cdot 10^7, \qquad &k_\mathcal{N} &= 2.5, \qquad &l_\mathcal{N} &= 1.25.
\end{align*}
The incorporation of the function ${h}$ accounts for varying incident flow directions. It is mainly the squared cosine of the angle of attack, which is regularized to ensure smoothness for tangential incident flow situations with regularization parameter $\epsilon = 10^{-7}$,
\begin{align*}
{h}(w_t, w) =
\begin{cases}
(w_t w^{-1})^2 , \qquad & w \geq \epsilon,\\
\left( 1- (w\epsilon^{-1})^2 \right)^2 + \left(3 - 2(w\epsilon^{-1})^2 \right)(w_t w\epsilon^{-2})^2, \qquad & w < \epsilon.
\end{cases}
\end{align*}

\setcounter{figure}{0}
\setcounter{equation}{0}
\section{Model for local turbulent velocity fluctuations} \label{appendix_turbRecon}
The turbulence reconstruction (cf.\ Sec.~\ref{subsubsec:turbRecon}) goes back to the works \cite{marheineke:p:2006, marheineke:p:2009b}. Given a $k_\star$-$\epsilon_\star$ description of the turbulent airflow,  the local turbulent velocity fluctuations are modeled as homogeneous isotropic incompressible Gaussian random field in space and time $\mathbf{v}_{\star,loc}'$ (with expectation $\mathbb{E}(\mathbf{v}_{\star, loc}')=\mathbf{0}$) that depend parametrically on the local kinematic viscosity and mean velocity of the airflow. Its covariance function is prescribed as product of the initial correlations transported with the local mean velocity and their temporal decay
\begin{align*}
\mathbb{E}(\mathbf{v}_{\star,loc}'(\mathbf{x}+\mathbf{y},t+s; \xi, \mathbf{w})\otimes\mathbf{v}_{\star,loc}'(\mathbf{x},t; \xi, \mathbf{w})) = \boldsymbol{\gamma}(\mathbf{y}-\mathbf{w}t;\xi)\,\Phi(s)
\end{align*}
with $\boldsymbol{\gamma}:\mathbb{R}^3\times\mathbb{R}\rightarrow\mathbb{R}^3$ isotropic and $\Phi:\mathbb{R}\rightarrow\mathbb{R}$ even for any $\mathbf{x}, \mathbf{y}, \mathbf{w}\in\mathbb{R}^3$ and $t,s,\xi\in\mathbb{R}^+_0$. Regarding the local kinetic turbulent energy and dissipation rate the following relations (in dimensionless formulation) hold
\begin{align*}
\frac{1}{2}\mathbb{E}(\mathbf{v}_{\star,loc}'(\mathbf{x},t; \xi, \mathbf{w})\cdot\mathbf{v}_{\star,loc}'(\mathbf{x},t; \xi, \mathbf{w})) &= 1, \qquad \mathbb{E}(\nabla_{\mathbf{x}}\mathbf{v}_{\star,loc}'(\mathbf{x},t; \xi, \mathbf{w}) : \nabla_{\mathbf{x}}\mathbf{v}_{\star,loc}'(\mathbf{x},t; \xi, \mathbf{w})) = \frac{1}{\xi},
\end{align*}
where $:$ denotes the scalar product for matrices and $\nabla_\mathbf{x}$ the nabla operator with respect to the variable $\mathbf{x}$. Due to the assumptions of isotropy and incompressibility the Fourier-transform $\mathcal{F}_{\boldsymbol{\gamma}}$ of the initial correlations $\boldsymbol{\gamma}$ can be expressed in terms of a single scalar-valued function, the so-called energy spectrum $E:\mathbb{R}^+_0\times\mathbb{R}^+_0\rightarrow\mathbb{R}^+_0$, i.e.,
\begin{align*}
\mathcal{F}_{\boldsymbol{\gamma}}(\boldsymbol{\kappa};\xi) = \frac{1}{4\pi}\frac{E(\lVert \boldsymbol{\kappa}\rVert;\xi)}{\lVert\boldsymbol{\kappa}\rVert^2}\left(\mathbf{I} - \frac{1}{\lVert\boldsymbol{\kappa}\rVert^2}\boldsymbol{\kappa}\otimes\boldsymbol{\kappa} \right)
\end{align*}
with unit matrix $\mathbf{I}\in\mathbb{R}^{3\times 3}$. Then, the above relations for turbulent energy and dissipation rate become
\begin{align}\label{sec:appB_eq:conditions_kappa}
\int_{\mathbb{R}^+_0} E(\kappa;\xi) d\kappa = 1,\qquad \int_{\mathbb{R}^+_0} \kappa^2 E(\kappa;\xi) d\kappa = \frac{1}{2\xi}.
\end{align}
In contrast to \cite{marheineke:p:2006, marheineke:p:2009b}, we use here a simplified energy spectrum 
\begin{align*}
E(\kappa;\xi) = \begin{cases}
\frac{1}{2}\kappa^{-5/3}, & \,\kappa_1\leq \kappa \leq \kappa_2, \qquad \quad \kappa_i=\kappa_i(\xi)\\
0, & \,\text{else},
\end{cases}
\end{align*}
which only reflects Kolmogorov's 5/3-Law but ensures the differentiability of the Gaussian field and enables the corresponding distribution function $F_E:[\kappa_1,\kappa_2]\times\mathbb{R}^+_0\rightarrow [0,1]$ and its inverse $F_E^{-1}:[0,1]\times\mathbb{R}^+_0\rightarrow[\kappa_1,\kappa_2]$ to be stated explicitly by analytic expressions
\begin{align*}
F_E(\kappa;\xi) = \frac{3}{4}\left(\kappa_1^{-2/3}-\kappa^{-2/3}\right),\qquad F_E^{-1}(y;\xi) = \left(\kappa_1^{-2/3} - \frac{4}{3}y\right)^{-3/2}.
\end{align*}
The $\xi$-dependent wave numbers $\kappa_i$, $(i=1,2)$, $0 < \kappa_1 < \kappa_2$ result from the relations in \eqref{sec:appB_eq:conditions_kappa} as
\begin{align*}
\kappa_i =\left(\frac{3}{2(\sqrt{y_E+1}+(-1)^{i+1})} \right)^{3/2}, \qquad y_E = \frac{1}{2}\left(b_E+\sqrt{\frac{2a_E}{b_E}-b_E^2}\right),
\end{align*}
with $a_E = (54/8\xi)^2$, $b_E = (c_E/18^{1/3} - (128/3)^{1/3}a_E/c_E)^{1/2}$, $c_E = (9a_E^2 + \sqrt{786a_E^3+81a_E^4})^{1/3}$.
The analytic expressions are advantageous in the sampling of the random field as they speed up drastically the numerical computations. Concerning the temporal correlation function we use the model of \cite{huebsch:p:2013, marheineke:p:2009b}
\begin{align*}
\Phi(t) = \exp\left(-\frac{t^2}{2t_L^2}\right)
\end{align*}
with $t_L = 0.212$ being the dimensionless decay time of the turbulent structures.
\setcounter{figure}{0}
\setcounter{equation}{0}
\section{Viscous stationary uni-axial fiber jet model and numerical scheme}\label{appendix_statModel}
For stationary considerations the fiber jet model is formulated in the Eulerian (spatial) description. This is achieved by re-parametrization of the Lagrangian (material) setting of System~\ref{system:Final} via an oriented time-dependent bijective mapping
\begin{align*}
S(\cdot,t): [-t,0]\rightarrow[S(-t,t),S(0,t)],\qquad \zeta\mapsto S(\zeta,t).
\end{align*}
Assuming sufficient regularity, a scalar convective speed $u$ and a spatial Jacobian $j$ corresponding to $S$ is defined by
\begin{align*}
\partial_tS(\zeta,t) = u(S(\zeta,t),t), \quad \partial_\zeta S(\zeta,t) = j(\zeta,t) > 0, \quad \text{with} \quad \partial_s u(S(\zeta,t),t) = \frac{\partial_t j}{j}(\zeta,t).
\end{align*}
In the Eulerian description the scalar speed $u$ becomes the Lagrangian multiplier to the assumption of global arc-length parametrization $\lVert\partial_s\mathbf{r}\rVert = 1$ and is hence an additional unknown of the system.

Starting from System~\ref{system:Final}, the stationary problem in Euler description for an uni-axial fiber with orientation $\boldsymbol{\tau} = \mathbf{e_g}$ and purely viscous material behavior ($\mathrm{De}\rightarrow0$) is given in non-dimensional form by System~\ref{system:Final_viscous} on the stationary domain $\Omega = (0,1)$ with tangential aerodynamic force $f_{air}$, i.e.\ $f_{air} = \mathbf{f}_{air}\cdot\boldsymbol{\tau}$.
The pressure of the viscous fiber jet model satisfies the relation $p = -{\sigma}/{3}$. 

\begin{system}[Stationary viscous uni-axial fiber model]\label{system:Final_viscous}
Kinematic equations as well as material laws in $\Omega$:
\begin{align*}
\frac{\mathrm{d}}{\mathrm{d}s} \sigma &= \frac{\mathrm{Re}}{3\mu}\sigma\left(u+\frac{\sigma}{u}\right) - \frac{1}{\mathrm{Fr}^2}u - f_{air}u,\\
\frac{\mathrm{d}}{\mathrm{d}s} T &= -\frac{\mathrm{St}}{\varepsilon}\pi d \alpha (T-T_\star),\\
\frac{\mathrm{d}}{\mathrm{d}s} u &= \frac{\mathrm{Re}}{3\mu}\sigma,
\end{align*}
Boundary conditions:
\begin{align*}
\sigma(1) = 0, \qquad T(0) = 1, \qquad u(0) = 1.
\end{align*}
\end{system}

System~\ref{system:Final_viscous} is a boundary value problem of ordinary differential equations of first order on a fixed domain. For its numerical solution we employ the continuation-collocation method, which has been used successfully in \cite{arne:p:2018, arne:p:2011, wieland:p:2018b}. The collocation method is a three-stage Lobatto IIIa formula, see, e.g., \cite{hairer:b:2009}. Mesh selection and error control are based on the residual of the continuous solution \cite{kierzenka:p:2008}. The resulting nonlinear system of equations is solved using a Newton method with numerically approximated Jacobian. This is a classical approach that is provided in the software MATLAB\footnote{For details see \texttt{http://www.mathworks.com}.} by the routine \texttt{bvp4c.m}. Its applicability depends on the convergence of the Newton method that is crucially determined by the initial guess. The initial guess is adapted iteratively by means of a continuation method from \cite{arne:p:2018, arne:p:2011, wieland:p:2018b}. For System~\ref{system:Final_viscous} we introduce a single continuation parameter $c \in[0,1]$ for the viscous, gravitational and aerodynamic forces as well as for the heat exchange. In particular we set
\begin{align*}
\frac{\mathrm{d}}{\mathrm{d}s} \sigma &= \frac{\mathrm{Re}}{3(c\mu+(1-c))}\sigma\left(u+\frac{\sigma}{u}\right) - c\frac{1}{\mathrm{Fr}^2}u - cf_{air}u,\\
\frac{\mathrm{d}}{\mathrm{d}s} T &= -c\frac{\mathrm{St}}{\varepsilon}\pi d\alpha(T-T_\star),\\
\frac{\mathrm{d}}{\mathrm{d}s} u &= \frac{\mathrm{Re}}{3(c\mu+(1-c))}\sigma.
\end{align*}
The starting solution for the continuation that belongs to $c=0$ is a stress-free straight fiber with constant speed and temperature taken from the inflow, i.e., $
\sigma=0$, $T=1$ and $u=1$.
%

\end{document}